\documentclass[titlepage,11pt,twocolumn]{article}

\usepackage{amsmath,amssymb,amsfonts,bbm}
\usepackage[linesnumbered,ruled,vlined]{algorithm2e}   % algorithms
\usepackage{booktabs,multirow}                         % better format in tables
\usepackage[authoryear]{natbib}                        % bibliography
\usepackage{tikz}                                      % for Fig. 1,3
\usepackage{forest}  
\usetikzlibrary{tikzmark}
\usepackage{todonotes}
\usepackage[pagewise,mathlines]{lineno}
\usepackage{setspace}

\newcommand{\ee}{\mathrm {e}}
\newcommand{\D}[1]{\mathop{\mathrm{d}{#1}}}
\newcommand{\EE}{\mathbb{E}}
\newcommand{\density}{h}
\newcommand{\forw}[1]{\overrightarrow{#1}}
\newcommand{\back}[1]{\overleftarrow{#1}}
\newcommand{\bafo}[1]{\overleftrightarrow{#1}}

\newcommand{\cont}[1]{{#1}^{c}}

\DeclareMathOperator{\Exp}{Exp}

\makeatletter
\renewcommand{\maketitle}{\bgroup\setlength{\parindent}{0pt}
\begin{flushleft}
  \LARGE\textbf{\@title} \\ \vspace*{2em}

  \large\@author\\ \vspace*{2em}
  
  \@date
\end{flushleft}\egroup
}
\makeatother

\bibpunct[, ]{(}{)}{;}{a}{}{;}

\title{Modelling and simulating Lenski's long-term evolution experiment}
\author{
     Ellen Baake$^{a,\ast}$, Adri\'{a}n Gonz\'{a}lez Casanova$^b$, 
     Sebastian Probst$^a$, Anton Wakolbinger$^{c,\ast}$ \\ \vspace*{1em}
     $^a$University of Bielefeld, Faculty of Technology, 33615 Bielefeld, Germany;  \\
     $^b$Universidad Nacional Aut\'{o}noma de M\'{e}xico (UNAM), Instituto de Matem\'{a}ticas, 04510 Ciudad Universitaria, M\'{e}xico;  \\
     $^c$Goethe University, Institute of Mathematics, 60629 Frankfurt am Main, Germany \\ \vspace*{1.5em}
     Email adresses: \\
     ebaake@techfak.uni-bielefeld.de (Ellen Baake)\\
     adriangcs@matem.unam.mx (Adri\'{a}n Gonz\'{a}lez Casanova)\\
     sprobst@techfak.uni-bielefeld.de (Sebastian Probst)\\
     wakolbinger@math.uni-frankfurt.de (Anton Wakolbinger)
}
\date{\textit{Please indicate}: authors in alphabetical order.}
\begin{document}

%%% coverpage 1
%
\pagenumbering{roman}

\maketitle

\newpage
%
% ---------------------------------
%
%%% coverpage 2
\subsubsection*{$^\ast$ Corresponding author}
 Ellen Baake, University of Bielefeld, Faculty of Technology, Universit\"{a}tsstra{\ss}e 25,  33615 Bielefeld, Germany, +49-521-106-4896, ebaake@techfak.uni-bielefeld.de \\[2mm]
\normalsize\noindent
\textbf{\textit{Keywords --- }} Lenski's long-term evolution experiment, epistasis, clonal interference, runtime effect,  Cannings model, offspring variance  \vspace*{1.5em} \\
\noindent
\textbf{\textit{Declaration of interest:}} none
\vspace*{1.5em} \\
\noindent
\textbf{\textit{Role of funding source:}}  Deutsche Forschungsgemeinschaft (German Research Foundation, DFG) provided financial support via Priority Programme SPP 1590 (Probabilistic Structures in Evolution, grants no. BA 2469/5-2 and WA 967/4-2), but was not   involved in the study design; in the collection, analysis and interpretation of data; in the writing of the report; or in the decision to submit the article for publication.
%
%
% ---------------------------------
%
%%% abstract
%
\begin{abstract}
\thispagestyle{plain}\setcounter{page}{2}
We revisit the model by Wiser, Ribeck, and Lenski (Science \textbf{342} (2013), 1364--1367), which describes how  the mean fitness increases over time due to beneficial mutations in Lenski's long-term evolution experiment. We develop the model further both conceptually and mathematically. Conceptually, we describe the experiment with the help of a Cannings model with mutation and selection, where the latter includes diminishing returns epistasis. The analysis sheds light on the growth dynamics within every single
day and reveals a runtime effect, that is, the shortening of the daily growth
period with increasing fitness; and it allows to clarify the contribution of epistasis to the mean fitness curve. Mathematically, we explain rigorous results in terms of a law of large numbers (in the limit of infinite population size and for a certain asymptotic parameter regime), and present approximations based on heuristics and supported by  simulations for finite populations.   
\end{abstract}
%
% ---------------------------------
%
%%% main content
%
\pagenumbering{arabic}\setcounter{page}{1}
\section{Introduction}
\label{sec:LTEEWRL}
One of the most famous instances in experimental evolution is Lenski's long-term evolution experiment or LTEE \citep{Lenski91,WRL13,Tenaillon16,Good17}. Over a period of 30 years, populations of \emph{Escherichia coli} maintained by daily serial transfer have  accumulated mutations, resulting in a steady increase in fitness. The mean fitness  is observed to be  a concave function of time, that is, fitness increases more slowly as time goes by. \citet{WRL13} formulated a  theoretical model that builds on the underlying processes, namely mutation, selection, and genetic drift, and obtained a good agreement with the data.  However, the model describes the underlying population processes  in a   heuristic way. As a consequence, one works with effective parameters that are hard  to interpret, and it is difficult to disentangle the contributions of the various model components  to the resulting fitness curve.  \citet{GKWY16} recently formulated an individual-based model for a special case (namely, for the case of deterministic fitness increments) and made explicit that the specific design of the LTEE lends itself ideally to a description via a \emph{Cannings model} \citep[Ch.~3.3]{Ewens04}. In a neutral setting, this classical model of population genetics works by assigning in each time step  to each of $N$ (potential) mothers indexed $j=1,\ldots, N$ a random number $\nu_j$ of daughters   such that the  $\nu_j$ add up to $N$ and are {\em exchangeable}, that is, they have a joint distribution that is invariant under permutations of the mother's indices. In  \citet{GKWY16}, this was  extended to include mutation and selection. 
While \citet{WRL13} work close to the data and perform an approximate analysis in the spirit of theoretical biology, \citet{GKWY16} focus on a precise definition of the model and on mathematical rigour (including in particular the proof of a law of large numbers in the infinite  population size limit and for a suitable parameter regime). 

The goal of this paper is to build a bridge between the two approaches,  to generalise the model of \citet{GKWY16} to random fitness increments, and to also consider it in the finite-population regime.
 A~thorough mathematical analysis will reveal the many connections between this model and the one of  \citet{WRL13}; in particular, this will make the meaning of its parameters transparent and will allow to separate the effects of the various model ingredients. Parameter identification and stochastic simulations of a suitable extension of the model will make the connection to the experimental data.
Let us briefly describe the LTEE  and the  outline of this paper.

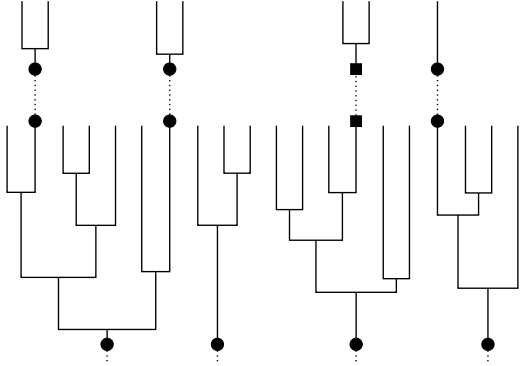
\begin{figure}
\begin{center}
\resizebox{.9\columnwidth}{!}{% Need to load packages 'tikz' and 'forest' in main file!!!
%
%\usepackage{tikz}
%\usepackage{forest}
%

\forestset{
  mytree/.style={
    for tree={
      edge+={thick},
      edge path'={
        (!u.parent anchor) -| (.child anchor)
      },
      grow=north,
      scale=0.7,
      parent anchor=children,
      child anchor=parent,
      anchor=base,
      l sep=5pt,
      s sep=10pt,
      if n children=0{align=center, base=bottom}{coordinate}
    }
  }
}

\begin{forest}
mytree
% right most tree
[,phantom, edge=dotted, 
 [, edge=dotted
  [, l = 10pt, edge=dotted, draw, circle, fill=black
    [, l=33pt
      [,tier=word]
      [, l=43pt
        [, l=13pt
          [,tier=word]
          [,tier=word]
        ]
        [,tier=word, draw, circle, fill=black
          [, edge=dotted, draw, circle, fill=black
            [, tier=word2]
          ]
        ]
      ]
    ]
  ]
 ]
% second right most tree 
 [, edge=dotted
  [, l = 10pt, edge=dotted, draw, circle, fill=black
    [
      [, l = 8pt
        [,l = 1pt, tier=word]
        [,l = 1pt, tier=word]
      ]
      [
        [, l = 28pt
          [,tier=word, draw, rectangle, minimum width=9pt, minimum height=9pt, fill=black
            [, edge=dotted, draw, rectangle, minimum width=9pt, minimum height=9pt, fill=black
              [, l = 15pt 
              	[, l = 1pt, tier=word2]
              	[, l = 1pt, tier=word2]
              ]
            ]
          ]
          [,tier=word]
        ]
        [, l = 18pt
          [,tier=word]
          [,tier=word]
        ]
      ]
    ]
  ]
 ]
% second left most tree
 [, edge=dotted
  [,l = 10pt, edge=dotted, draw, circle, fill=black
    [, l=70pt
      [
        [,tier=word]
        [,tier=word]
      ]
      [,tier=word]
    ]
  ]
 ]
%left most tree
 [, edge=dotted
  [,l = 10pt, edge=dotted, draw, circle, fill=black
    [, l = 4pt
      [, l = 34pt
        [,tier=word, draw, circle, fill=black
          [, edge=dotted, draw, circle, fill=black
            [, l = 1pt 
             [, tier=word2]
             [, tier=word2]
            ]
          ]
        ]
        [,tier=word]
      ]
      [
        [
          [,tier=word]
          [
            [,tier=word]
            [,tier=word]
          ]
        ]
        [, l = 50pt
          [,tier=word, draw, circle, fill=black
            [, edge=dotted, draw, circle, fill=black
              [, l = 12pt 
              	[,tier=word2]
              	[,tier=word2]
              ]
            ]
          ]
          [,tier=word]
        ]
      ]
    ]
  ]
 ]
]
\end{forest}}
\end{center}
\caption{Illustration of some day $i - 1$ (and the beginning of day $i$) of Lenski's LTEE with $4$ founder individuals (bullets), their  offspring trees within day $i-1$, and the  sampling from day $i-1$ to $i$ (dotted), for an average clone size of $5$. The second founder from the left at day \mbox{$i - 1$} (and its offspring) is lost due to the sampling, and the second founder from the right at day $i$ carries a new beneficial mutation (indicated by the square). }
\label{fig:forest}
\end{figure}

\paragraph{Lenski's LTEE.} Every morning, Lenski's LTEE starts with a sample of $\approx~5 \cdot 10^6$ \emph{Escherichia~coli} bacteria in a defined amount of fresh minimal glucose medium. During the day (possibly after a lag phase), the bacteria divide until the nutrients are used up; this is the case when the population has reached $\approx 100$ times its original size. The cells then  stop dividing and enter a stationary phase. At the end of the growth period, there are therefore $\approx 5 \cdot 10^8$ bacteria, namely, $\approx 5 \cdot 10^6$ clones each of average size $\approx 100$, see Fig.~\ref{fig:forest}.   The next morning, one takes a random sample of $\approx 5 \cdot 10^6$ out of the $\approx 5 \cdot 10^8$ cells, puts them into fresh medium, and the process is repeated;  the sampled individuals are the roots of the new offspring trees. Note that the number of offspring a founder individual contributes to the next 
day is random; it is 1 on average, but can also be $0$ or greater than one.

Lenski started 12 replicates of the experiment in 1988, and since then it has been running without interruption. The goal of the experiment is to observe evolution in real time. Indeed, the bacteria evolve  via beneficial mutations, which allow them to adapt to the environment and thus to  reproduce faster. 

\begin{figure*}[!ht]\centering
\resizebox{.63\textwidth}{!}{
\input{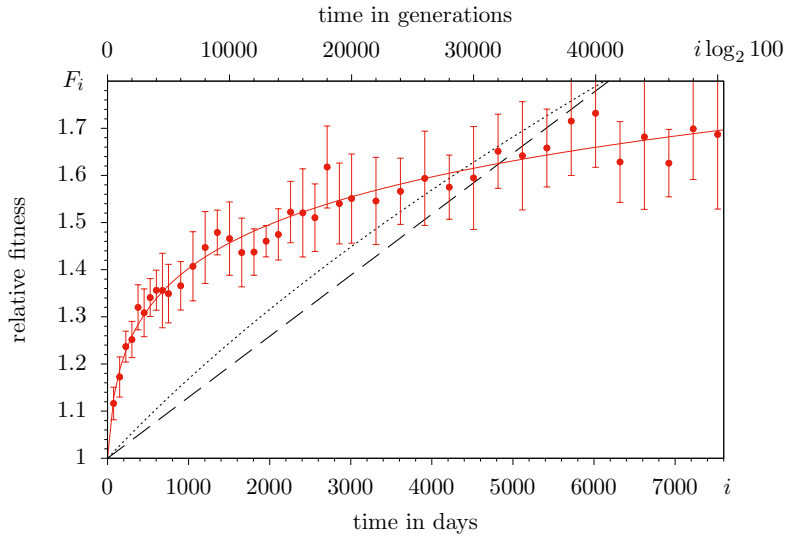}
}
\caption{Empirical relative fitness averaged over all 12  populations (red bullets) with error bars (95\% confidence limits based on the 12 populations) from \cite{WRL13};
%and averaged over the 6  populations (black triangles) that kept the original (low) mutation probability (that is, hypermutator strains were %excluded); 
and corresponding power law \eqref{powerlaw} with $\widehat{g} = 5.2$ and $\widehat{\beta}=5.1 \cdot 10^{-3}$ (red solid line).
% and $\widehat{g} = 6.0$, $\widehat{\beta}=8.7 \cdot 10^{-3}$ (black dashed line), respectively. 
Data and parameters according to Fig.~2A and Table~S4 of \citet{WRL13DATA}. The best fit of a square root (black, dotted) and a linear (black, dashed) fitness trajectory is also shown; these correspond to a scenario without epistasis  with and without runtime effect, respectively, as explained in Sec.~ \ref{sec:GKWYLLN}.}
\label{fig:WRL_data}
\end{figure*}

One special feature of the LTEE is that samples are frozen at regular intervals. They can be brought back to life at any time for the purpose of comparison  and thus form a living fossil record. In particular, one can, at any day~$i$, compare the current population with the initial (day $0$) population via the following \emph{competition experiment} \citep{LT94,WRL13}. A  sample from the day-0 population and one from the day-$i$ population, each of the same size, are grown  together; we define $T_i$ as the time at which the nutrients are used up.  One then defines the \emph{empirical relative fitness at day} $i$ as
\begin{equation}\label{emp_rel_fitness}
\widetilde F_i = \frac{\log \big ( Y_i(T_i)  /  Y_i(0) \big )}{\log \big ( Y_0(T_i)  /  Y_0(0) \big )}, 
\end{equation}
where, for $T=0$ and $T=T_i$,  $Y_i(T)$ and $Y_0(T)$ are the sizes at time $T$ of the populations grown from the day-$i$ sample and the day-$0$ sample, respectively. Note that the empirical relative fitness is a random quantity, whose outcome will vary from replicate to replicate. Fig.~\ref{fig:WRL_data}  shows the time course over 21 years of the empirical relative fitness averaged over the  replicate populations,   as reported by \cite{WRL13}. Obviously, the \emph{mean relative fitness} has a tendency to increase, but the increase levels off, which leads to a conspicuous concave shape.

% henceforth referred to as WRL
% Both epistasis and clonal interference may be considered as interactions of mutations (within one individual in the case of epistasis, between different individuals in the case of clonal interference). 

As noted by \citet{WRL13}, the mean relative fitness   may be described by
 the  power law
\begin{equation}\label{powerlaw}
{\widetilde f} \big( k  \,\big) = \big(1+\beta k \,\big)^{\frac{1}{2 g}}
\end{equation}
with parameters $\beta> 0$ and $g >0$. Here $\beta$ is a time-scaling constant, and the exponent $g$ determines the shape of the curve. Furthermore, $k$ is time with one generation (which  here is the mean doubling time) as unit, so
\begin{equation}\label{t_gamma}
 i = \Big \lfloor \frac{k}{\log_2 100} \Big \rfloor \approx \frac{k}{6.6}\,.
\end{equation}
The red solid line in Fig.~\ref{fig:WRL_data} shows the best fit of this curve to the data of all 12 replicate populations, as obtained by \cite{WRL13}, with parameter estimates 
%$\widehat g = 5.26$
 $\widehat g = 5.2$ and $\widehat \beta = 5.1 \cdot 10^{-3}$. (Here and in what follows,  parameter values estimated from the data are indicated by a hat, and numbers are rounded to 2 digits.  Our parameters obtained via \texttt{NonlinearModelFit} of \texttt{Wolfram Mathematica 11} only differ in the third digits.) 
 %the black dashed curve is the corresponding fit after exclusion of those populations that evolved into hypermutator strains. 
  In line with  \eqref{emp_rel_fitness} and \eqref{t_gamma}, we take \emph{days} as our discrete time units, rather than doubling times (this will pay off in Secs.~\ref{sec:GKWYLLN} and~\ref{sec:ci}); so $\log_2 100 \approx 6.6$ generations in Fig.~\ref{fig:WRL_data} correspond to one day, and the total of 50000 generations correspond to around 7525 days. 
 
The two models mentioned above  aim to explain the power law \eqref{powerlaw}. The one by \cite{WRL13}, which we will refer to as the WRL model,   uses  an approach of \emph{diminishing returns epistasis}, which means  that the beneficial effect of mutations decreases with increasing fitness (cf.\ \citet[p.~74]{Bu00} or \cite{Petal00}). \cite{WRL13} derive, by partly heuristic methods, a differential equation for the mean relative fitness whose solution is given by \eqref{powerlaw}. The time-scaling parameter $\beta$ is determined by the interplay of the rate and  the effect of beneficial mutations, with the heuristics of \cite{GL98} for the description of clonal interference playing an important role.\footnote{\label{foot_CI}Clonal interference \citep{GL98,Ge01,PK07} refers to the situation of two (or more) beneficial mutations present in the population at the same time. They then compete with each other and, in the end, only one of them will be established in the population; an effect that slows down adaption (when measured against the stream of incoming mutations), and biases the distribution of beneficial effects.}

The second approach is the individual-based model of \cite{GKWY16} and makes full use of  ideas, concepts, and techniques from mathematical  population genetics, which seem to be ideally tailored for the LTEE setup. We will address this as the GKWY model; since it has been published  in a mathematical journal, we will review it in more detail in Sec.~\ref{sec:GKWYLLN} with an emphasis on the biological content.  For a certain parameter regime that excludes clonal interference, and using a similar approach to diminishing returns  as in the WRL model, \citet{GKWY16} prove a law of large numbers as $N\to \infty$, thereby rigoroulsy deriving a version of the power law~\eqref{powerlaw}. 

\paragraph{Goal and outline of this paper.}
A major goal of this paper is to provide a thoroughly-founded mathematical model of the LTEE,  and to relate it to the observed fitness curve via parameter estimation and stochastic simulations. This approach will provide additional connections between the ideas contained in the WRL and the GKWY models addressed in the previous paragraph.
%
%We will build on the recent paper by \citet{GKWY16} in the mathematical literature and partially lean on the WRL model, which is the standard in theoretical biology.
The  design of the LTEE, with  the daily  growth cycles and the sampling scheme, results in an (approximately) constant population size at the beginning of each day.   
 As made explicit by \citet{GKWY16}, this  lends itself in a prominent way to a description through a Cannings model (including mutation and selection), where the mothers are identified with the founders in a given day and the daughters with the founders in the next day. 
  The crucial parameter of the Cannings model, namely, the \emph{variance of the number of offspring} of a founder individual that make it into the next day, is obtained in the context of the LTEE from an explicit stochastic model of population growth during each day. This offspring variance enters Haldane's formula for the fixation probability, see  \eqref{Haldane} below. 
  
  As a matter of fact, also \citet{WRL13} use  a  formula for the fixation probability (see Eq. (S1) in their Supplementary Text). In this context they refer to \cite{GL98}, who assume a deterministic population growth (and clones of equal size) resulting from synchronous divisions. Indeed, the Cannings model  thus  hidden within the WRL model turns out to work with a different offspring variance; we will come back to this in Sec.~\ref{sec:discussion}.
  
%  Between the lines, however, a Cannings model is already hidden in the WRL model; but here deterministic population growth resulting from synchronous divisions is assumed implcitly \citep{GL98,WRL13} and results in a different offspring variance and hence in a different version of the Cannings model. In an equivalent way, the notions of \emph{pair coalescence probability} and \emph{effective population size} rely on the intraday model of population growth. 
 
In addition to the specification of the offspring variance, our model for the daily population growth in continuous time allows us to quantify selection (including diminishing returns epistasis) at the level of the individual reproduction rates within a day.  The effect of diminishing returns seems to be obvious from Fig.~\ref{fig:WRL_data}; however, epistasis is not the only contribution to the fitness curve. Rather, the design of the experiment also has its share in it via what we call the \emph{runtime effect}, namely, the shortening of the daily growth phase with increasing fitness. In fact, the runtime effect \emph{alone} results in a concave fitness curve, but is not strong enough to explain the observed data in the absence of epistasis. The analysis of our model will allow a clear separation of the respective contributions. Likewise, the  population-genetic notions that also appear in the WRL model (namely, the mutation rate, the selective advantage, the effective population size, the fixation probability, and the strength of epistasis) will be made precise in terms of the underlying microscopic model. Throughout, we aim at a rigorous mathematical treatment where possible.

The paper is organised as follows. In Sec.~\ref{sec:GKWYLLN}, we will recapitulate the GKWY model  and explain its law of large numbers (that is a deterministic limit in a suitable parameter regime as population size goes to infinity) for a more biological readership.  At the end of Sec.~\ref{sec:GKWYLLN}, we will consider  the resulting stochastic effects in a system whose parameters are obtained from a fit to the data observed in the LTEE (and which thus naturally differs from its infinite population limit). In Sec.~\ref{sec:ci}, this will lead us to consider 
clonal interference, which we will investigate  both for the case of deterministic and 
 random fitness increments. Here we do not prove a law of large numbers, but derive approximations with the help of moment closure and a refined version of the Gerrish-Lenski heuristics.
In Sec.~\ref{sec:discussion}, we will thoroughly discuss the crucial  differences between the WRL and the GKWY models, together with the key notions of fitness increment, selective advantage,   and epistasis, as well as the mutually equivalent concepts of offspring variance, pair coalescence probability, and effective population size.
\section{A probabilistic model for the LTEE and a law of large numbers}
\label{sec:GKWYLLN}
The GKWY model  takes into account two different dynamics, namely, the dynamics \emph{within each individual day}, and the dynamics \emph{from day to day}, together with a suitable \emph{scaling regime}. The resulting \emph{relative fitness process} is proved to converge, in the $N \to \infty$ limit, to a power law equivalent to \eqref{powerlaw}; that is, the power law arises as a \emph{law of large numbers}. We explain this here with the help of  an appropriate \emph{heuristics}. In what follows, we present these building blocks and perform a first \emph{reality check}.

\paragraph{Intraday dynamics.} Let $T$ be (continuous) physical time  within a day, with $T=0$ corresponding to the beginning of the  growth phase (that is, we discount the lag phase). Day $i$ starts with
$N$  founder individuals ($N \approx 5 \cdot 10^6$ in the experiment). The reproduction rate or \emph{Malthusian fitness}\footnote{As in the WRL model, the simplifying assumption is inherent here that fitness is equivalent to  reproduction rate, whereas other phenomena may also influence the composition of the final population  from which one samples to seed the next-day’s culture; such as the duration of the lag phase  or the ability to sustain some growth even when nutrients have become sparse.}  of founder individual~$j$ at day $i$ is 
$R_{ij}^{}$, where $i \geqslant 0$ and $1  \leqslant j \leqslant N$. It is assumed that at day 0   all individuals have identical rates,  $R_{0j}^{} \equiv R_0^{}$, so the population is \emph{homogeneous}. Offspring inherit the reproduction rates from their parents.

We use dimensionless variables right away. Therefore we denote by
\begin{linenomath}\postdisplaypenalty=0
\begin{equation}
	t = R_0 T	\quad \text{and } r^{}_{ij}  = \frac{R_{ij}}{R_0} \label{t} 
\end{equation}
\end{linenomath}
dimensionless time and rates, so that on the time scale $t$ there is, on average, one split per time unit at the beginning of the experiment (this unit is  $\approx 55$ minutes, cf. \citet{Barrick09}) and  $r^{}_{0j} \equiv 1$. In this paragraph, we consider the $r_{ij}^{}$ as given (non-random) numbers.

We thus have  $N$ independent \emph{Yule processes}  at day $i$:  all descendants of founder individual $j$ (the members of the $j$-clone) branch at rate $r^{}_{ij}$, independently of each other. They do so until $t=\sigma_i$, where
$\sigma_i$ is the duration of the growth phase  on day $i$. We define $\sigma_i$ as the value of $t$ that satisfies
\begin{equation}\label{def_sigma}
\begin{split}
& \EE(\text{population size at time } t) \\
& = \sum_{j=1}^N \ee^{r^{}_{ij}t} =\gamma N, 
\end{split}
\end{equation}
where $\gamma$ is, equivalently, the multiplication factor of the population within a day, the average clone size,  and the dilution factor from day to day in the experiment ($\gamma \approx 100$ in the LTEE). Note that the Yule processes are  stochastic, so the population size at time $t$ is, in fact, random; in the definition of $\sigma_i$, we have idealised by replacing this random quantity by its expectation. Since $N$ is very large, this is well justified, because the fluctuations of the random time needed to grow to a factor 100 in size are small relative to its expectation. Note also that the assumption of a fixed dilution factor implies the supposition of a one-to-one correspondence between the amount of nutrient and the population size at the end of the day; this may be violated if the cell size evolves over time, as was observed in some lines of the experiment \citep{LT94}.

\paragraph{Interday dynamics.} At the beginning of day $i > 0$, one samples $N$ new founder individuals out of the $\gamma N$ cells from the population at the end of day $i-1$. We assume that one of these new founders carries a \emph{beneficial mutation} with probability $\mu$; otherwise (with probability $1 - \mu$), there is no beneficial mutation. We think of $\mu$ as the probability that a beneficial mutation occurs in the course of day $i-1$ and  is sampled for day $i$. In the light of the constant number of cell divisions per day (regardless of the current fitness value), it is implied that  $\mu$ is independent of $i$. 

Assume that the new beneficial mutation at day $i$ appears in individual $m$, and that the reproduction rate of the corresponding founder individual $k$ in the morning of day $i-1$ has been  $r_{i-1, k}$. The new mutant's reproduction rate is then assumed to be
\begin{equation}\label{r_increase}
	r^{}_{im} = r^{}_{i-1,k} + \delta(r_{i-1,k}) \text{ with } \delta(r) := \frac{\varphi}{ r^{q}}.
\end{equation}
Here, $\varphi$ is the  beneficial effect due to the first mutation (that is $\delta(1)$, which applies while \mbox{$r=1$}),  and $q$ determines the strength  of epistasis. In particular, 
$q=0$ implies constant increments (that is, additive fitness), whereas
$q>0$ means that the increment decreases with $r$, that is, we have diminishing returns epistasis. 

Let us, at this point, include some comments on the modelling of both  fitness increments  and mutation.  As to fitness, note first that we only take into account beneficial mutations. While
neutral and deleterious mutations  are, in general, considered more frequent than beneficial ones \citep{EWK07},  we will follow \cite{WRL13} and work in a parameter regime of weak mutation and moderate selection, where beneficial mutations originate and go to fixation one by one, while neutral and deleterious mutations do not contribute to the  fitness trajectory \citep{McCS14}; in contrast to strong-mutation regimes that may lead to fitness waves including all kinds of mutations (as discussed, for example,  by \cite{MR16}). We also adhere to the simplistic assumption that the fitness landscape is \emph{permutation invariant}, that is, every beneficial mutation on the same background conveys the same deterministic fitness increment, no matter where it appears in the genome. This  simplification is abundant both in the classical (Fisher's (1918) \emph{staircase model}) and in the modern literature  \citep{DeFi07}. In particular, this entails that the effect of neutral networks is neglected. In fact,  neutral mutations can play important roles because they can explore distant fitness peaks via neutral networks  (see \cite{HSF96} for early work in the context of RNA structures, \cite{KCGP06} for an application to the evolution of influenza viruses, and \cite{MC15} for a recent study of the effects on the molecular clock).  The assumption will be relaxed in Sec.~\ref{sec:stoch}, where we turn to stochastic increments. 

As to the mutation model, let $M_i^{}$ be the number of new mutants in the sample of size $N$ at the beginning of  day $i$. So far we have assumed that $M_i$ can only take the values 1 or 0. More generally,  for  describing the random number of individuals  that are offspring of new mutants from day $i-1$ {\em and} make it into the $N$-sample at the beginning of day $i$, we might consider integer-valued random variables $M_i$  with small expectation $\mu$. The above definition of the mutation mechanism  means in particular that the mutation probability does not depend on the current fitness value. We keep this assumption also for the distribution of $M_{i}$, and, as in \eqref{r_increase}, suppose that any mutation adds $\delta(r)= \varphi /r^q$ to the pre-mutant reproduction rate. Unless  $\mu$ is  very small, realism may be added by using Poisson random variables, which is what we do in the simulations, see Appendix~C. One might also think of a finer \emph{intraday modelling} of the mutation mechanism, cf. \cite{WGSV02}, \cite{WZ15}, or \cite{LCW18}. Although the limit theorem in \cite{GKWY16} is proved only for binary random variables $M_i^{}$, we conjecture that its assertion also holds for non-binary $M_i$ in the scaling regime \eqref{scaling} discussed below, at least as long as the variances of the $M_i$ remain bounded as $N\to \infty$. We will adhere to the binary assumption in our analysis, and it will  lead to very satisfactory approximations, see Sec.~\ref{sec:ci}. Note also that we have idealised  by not taking into account the change in fitness due to mutation during the day; this is because a mutant appearing during the day will not rise to appreciable frequency in the course of this first day of its existence, and thus will not change the overall growth rate of the population in any meaningful way.
%\\\\
%As long as $\mu_N$ is not very small, precision may be added by using Poisson random variables, which is what we do in the simulations, see Appendix. One might also think of an even finer `intraday modelling' of the mutation mechanism, cf. \cite{WZ15} or \cite{LCW18}. H             owever, because of the design of the LTEE ($N \approx 5\cdot 10^6, \gamma \approx 100$), it does not seem likely that this should lead to substantial clustering phenomena (or even heavy tails) in the distribution of the number of new beneficial mutants in the next morning.
\paragraph{Mean relative fitness.} With a view towards~\eqref{emp_rel_fitness} we define the {\em mean relative fitness}, depending on the configuration of reproduction rates $r_{ij}$ of the $N$ individuals in the sample at the beginning of day $i$, as
\begin{equation}\label{rel_fitness}   
 F_i   :=    \frac{1}{\sigma_i} \log \Big ( \frac{1}{N} \sum_{j=1}^{N} \ee^{r^{}_{ij}  \sigma_i^{}} \Big ).
\end{equation}
Here, $\sigma_i$ is as defined in~\eqref{def_sigma}, so that $F_i = (\log \gamma)/\sigma_i.$
Comparing \eqref{emp_rel_fitness} and \eqref{rel_fitness}  we see that the former contains additional sources of randomness: on the one hand, the numerator of \eqref{emp_rel_fitness} may be viewed as stemming from a sample that was drawn from the population at the end of day $i-1$ (and which consists of individuals different from those present at the beginning of day $i$), on the other hand the duration of the growth phase leading to   \eqref{emp_rel_fitness} is not a predicted time as in \eqref{rel_fitness} but an empirical time coming out of the competition experiment between the samples from day  $i$ and day~$0$. However, since the samples consist of a large number of individuals, the random variables occurring in  \eqref{emp_rel_fitness} will, with high probability, come out close to their expectations, thus making already a single copy of the random variable \eqref{emp_rel_fitness} a reasonably good approximation of~\eqref{rel_fitness}, at least if the population  at day $i$ is sufficiently homogeneous. To see this, we take care of the new time scale and rewrite $\widetilde \sigma_i = R_{0} \, T_i$ together with $y_i(t) = Y_i(T)$ for $T=0$ and $T=T_i$. Assume that the competition experiment starts with a sample of size $y_0=n$ from the ancestral population and a sample of size $y_i=n$ from the day-$i$ population. In the `deterministic approximation' mentioned above, the duration of the experiment,  $\widetilde \sigma_i$, then is the solution of
\[
n\, \ee^t + \sum_{j=1}^n \ee^{r^{}_{ij} t} = 2\, n\, \gamma,
\]
so
\[
 y^{}_0(\widetilde \sigma^{}_i) \approx n \, \ee^{\widetilde \sigma^{}_i} \quad \text{and } 
  y^{}_0(\widetilde \sigma^{}_i) \approx \sum_{j=1}^n \ee^{r^{}_{ij} \widetilde \sigma^{}_i}.
\]
Consequently,
\begin{equation}\label{hatFi}
\widetilde{F}_i = \frac{\log \big ( y^{}_i(\widetilde \sigma^{}_i) /  y^{}_i(0) \big ) }{\log \big ( y^{}_0(\widetilde \sigma^{}_i) /  y^{}_{0}(0) \big )}
\approx \frac{1}{\widetilde \sigma^{}_i} \log \bigg ( \frac{1}{n} \sum_{j=1}^n \ee^{r^{}_{ij} \widetilde \sigma^{}_i} \bigg ).
\end{equation}
Due to the enhanced reproduction rates at day $i$ compared to day $0$, $\widetilde \sigma^{}_i$
will generically be larger than $\sigma_i$. This is because $\widetilde \sigma^{}_i$ refers to a mixture of day-$i$ and day-0 populations, 
whereas $\sigma_i$ relates to a `pure' day-$i$ population. But if the day-$i$ population is homogeneous, that is, $r_{ij} \equiv r_i$, 
one has $y^{}_i(\widetilde \sigma^{}_i) = n \, \ee^{r_i \widetilde \sigma_i}$, so $\widetilde \sigma_i$  cancels out in \eqref{hatFi}, and 
$\widetilde F_i \approx F_i$. If the population is  inhomogeneous,  however,  $\widetilde F_i $ will be systematically  larger than $F_i$, because then the individuals with a larger reproduction rate will get more weight in  \eqref{emp_rel_fitness} than in~\eqref{rel_fitness}.\label{fn:compare:emp_rel_fit:rel_fitness} Fortunately, it will turn out in Sect.~\ref{sec:stoch} that  polymorphism is low in our populations, so  $F_i$ may be taken as a valid approximation to $\widetilde F_i$.

%The obvious notion of mean relative fitness is the expectation of the empirical relative fitness in \eqref{emp_rel_fitness}, that is, 
%\begin{equation}\label{not_rel_fitness}
%  \EE \Bigg ( \frac{\log \big ( Y_i(T_i)  /  Y_i(0) \big )}{\log \big ( Y_0(T_i)  /  Y_0(0) \big )} \Bigg
%   )\,.
%\end{equation}
%For the sake of mathematical tractability, however, we instead  define the mean relative fitness at day $i$ as
%\begin{equation}\label{rel_fitness_def}
%F_i^N   := \frac{\log \Big ( \EE \big ( Y_i( T_i) \big ) / Y_i(0) \Big )}
%          {\log \Big ( \EE \big ( Y_0( T_i) \big ) / Y_0(0) \Big ) }.
%\end{equation}
%This is a valid approximation to \eqref{not_rel_fitness} for a reason similar to that given below \eqref{def_sigma}: For large $Y_i(0)$, the fluctuations of $Y_i( T_i)  / Y_i(0)$ are small relative to their expectation, so  the moment closure underlying the use of \eqref{rel_fitness_def} instead of \eqref{not_rel_fitness} seems justified.
% Assuming that $Y_i(0)=N$  and setting    $t_i = R_0 T_i$, where  $T_i$ is again the time until the nutrients are used up, \eqref{rel_fitness_def} turns into
%% \begin{equation}\label{rel_fitness}   
%% F_i^N   =    \frac{1}{t_i} \log \Big ( \frac{1}{N} \sum_{j=1}^{N} \ee^{r^{}_{ij} t_i^{}} \Big ) 
%%\end{equation}
%% (recall that the LTEE itself works with sample size $N$, but this size is divided between the two samples in the competition).
%since $r_{0j}^{} \equiv 1$, so $\EE \big ( Y_0( T_i) \big ) / Y_0(0)=\ee^{t_i}

Note  that
\eqref{rel_fitness} implies that
\begin{equation}\label{exp_rel_fitness}
 \ee^{F_i^{} \sigma_i^{}}   =   \frac{1}{N} \sum_{j=1}^{N} \ee^{r^{}_{ij} \sigma_i^{}}.  
\end{equation}
Thus,  $F_i$ may be understood as the \emph{effective reproduction rate} of the population at day $i$, which is different from the mean Malthusian fitness $\frac{1}{N} \sum_j r^{}_{ij}$ unless the population is homogeneous.

\paragraph{Heuristics leading to the limit law.} 
Assume a new mutation arrives in a \emph{homogeneous} population of relative fitness $F$. 
It conveys to the mutant individual a relative \emph{fitness increment} 
\begin{equation}\label{delta_N}
\delta (F) = \frac{\varphi}{F^q},
\end{equation}
that is, the mutant has relative Malthusian fitness $F+\delta (F)$.
The length of the growth period  then is 
\begin{equation}\label{sigma}
\sigma(F) =\frac{\log \gamma}{F}
\end{equation}
(since this solves $\ee^{Ft}=\gamma$, cf. \eqref{def_sigma}). We now define the \emph{selective advantage} of the mutant as 
\begin{equation}\label{s_N}
s(F) = \delta(F) \, \sigma(F).
\end{equation}
Obviously, \emph{the length $\sigma$ of the growth period decreases with increasing} $F$ and, since $s$ in \eqref{s_N} decreases with decreasing $\sigma$,  $s$ would decrease with increasing $F$ even if $\delta(F)$ were constant. This is what we call the
\emph{runtime effect:} adding a constant to an interest rate $F$ of a savings account becomes less efficient when the runtime decreases. 

Let us explain the reasoning behind \eqref{s_N}. In population genetics, the selective advantage (of a mutant over a wildtype) per generation is 
\begin{equation}\label{s}
s = \frac{a_1^{}-a_0^{}}{a_0^{}},
\end{equation}
where  $a_0^{}$ ($a_1^{}$) is the expected number of descendants of a wildtype (mutant) individual in one generation; Eq.~\eqref{s} has the form of a \emph{return} (of a savings account, say). If growth is in continuous time with Malthusian parameters $r_0^{}$ and $r_1^{}=r_0^{} + \delta$, respectively, and a generation takes time $\sigma$, then $a_0^{}=e^{r_0^{}\sigma}$ and  $a_1^{}=e^{r_1^{}\sigma} \approx a_0^{}\, (1 + \delta\,\sigma)$ if $\delta$ is small, which turns \eqref{s} into \eqref{s_N}.   Often, the appropriate notion of a generation is the time until the population has doubled in size, see e.g. Eq.~(3.2) in \citet{Chevin11}, which provides an analogue to \eqref{s_N}. In our setting, the corresponding quantity is the time required for the population to grow to $\gamma$ times its original size, which is the length $\sigma(F)$ of the growth period in \eqref{sigma}.\footnote{In line with this,  we choose days as our discrete time units, as already mentioned in Sec.~\ref{sec:LTEEWRL}.} Together with the above expression for $s$, this explains \eqref{s_N}.
Notably, a formula that is perfectly analogous to \eqref{s_N} also appears in \citet[p.~1977, last line]{Sanjuan10}; there, the concept of a viral generation is associated with the cell infection cycle, and the number $K$ (which corresponds to our~$\gamma$) is the burst size or viral yield per cell.

Furthermore, it is precisely this notion of selection advantage conveyed by \eqref{s_N} and \eqref{s}  that governs the \emph{fixation probability}. Namely, the fixation probability of the mutant turns out to be 
\begin{equation}\label{pi_N}
\pi(F) \sim C\, s(F).
\end{equation}
Here, $\sim$ means asymptotic equality in the limit $N \to \infty$, \footnote{That is, $\pi(F) / (C \, s(F)) = \pi^{}_N(F) / (C \, s^{}_N(F)) \to 1$ as \mbox{$N \to \infty$}, in the next paragraph’s setting; see \cite{GKWY16}.} and $C := \gamma/(\gamma - 1)$ is asymptotically twice the reciprocal offspring variance in one Cannings generation of the  GKWY model\footnote{Let us emphasise once again that one generation of this Cannings model corresponds to one day in the LTEE.}; that is, with the notation introduced in the first paragraph of the Introduction,  the offspring variance $v$ in one Cannings generation satisfies 
\begin{equation} 
\label{ourv}
v = \mathbb{V}(\nu_1) \sim 2 \, \frac{\gamma-1}{\gamma} = \frac{2}{C}.
\end{equation}
 Hence \eqref{pi_N} is in line with Haldane's formula
\begin{equation}\label{Haldane}
\pi \sim \frac s{v/2},
\end{equation}
which says that the fixation probability $\pi$ is (asymptotically) the selective advantage $s$ divided by half  the offspring variance $v$  in one generation. Haldane's formula relies on a branching process approximation of the initial phase of the mutant growth; see \citet{PW08} for an account of this method, including a historic overview. We will also encounter the branching approximation in  the argument around \eqref{branching} below.

For the sake of completeness, let us give the following intuitive explanation for \eqref{ourv}.
In every Cannings model, one has the
 relation
 \begin{equation} \label{vc}
v = (N-1) \, p_{\rm coal}
\end{equation}
between  $v$ and the pair coalescence probability $p_{\rm coal}$, that is, the probability that two randomly sampled daughters have the same mother, cf.~\citet[Ch.~4.1]{Dur08}. Eq.~\eqref{vc} then follows readily from the elementary relation $p_{\rm coal}= \mathbb E[\frac{1}{N \, (N-1)} \sum_j \nu_j \, (\nu_j-1)]$, which, in turn, equals $\frac {1}{N-1} (\mathbb E[\nu_1^2] -1)= \frac{1}{N-1} \, v$, because the $\nu_j$ are exchangeable and sum to $N$ by assumption.
In our specific Cannings model, the  family size of a randomly sampled daughter individual at the end of the day is, on average, asymptotically twice as large as a typical family size\footnote{The size of the clone to which a sampled individual  belongs has a {\em size-biased} distribution; this is in line with the classical  {\em waiting time paradox} (cf.~ \citet[Example~4.16]{HOG13}).  In our model, the  size distribution of a typical clone at the end of the day is approximately geometric with parameter $1/\gamma$, and the size-biasing of this distribution results  (approximately) in a negative binomial with parameters $2$ and $1/\gamma$.  Consequently, the expected size of the clone to which a sampled individual  belongs is approximately $2 \, \gamma$, that is twice  the expected size of a typical clone.  This proportion carries over from the clones to the families of sampled individuals. Let us emphasise once again that a \emph{family} consists of the founders at the beginning of the next day that go back to the same founder in the current day; whereas a \emph{clone} consists of all descendants of a founder at the end of a day, regardless of whether they are sampled for the next day or not.}.  Since we have $N$ clones of average size~$\gamma$, and the sampling is without replacement, we have 
\begin{equation}\label{ourPCP}
p_{\rm coal} \sim \frac {2}{N}  \, \frac{\gamma-1}{\gamma}.
\end{equation}
Together with \eqref{vc} this implies \eqref{ourv}. Note that \eqref{ourPCP}, at the same time, defines the (coalescence) effective population size via $N_{\text e}=1/p_{\rm coal}$, cf.~\citet[Ch.~3.7]{Ewens04} or \citet[Ch.~4.4]{Dur08}.

Another crucial ingredient of the heuristics is the time window of length 
\begin{equation}\label{u_N}
u(F) \sim \frac{\log \big (N \, s(F) \big )}{s(F)} 
\end{equation}
after the appearance of a beneficial mutation that will survive drift (a so-called \emph{contending mutation}); this approximates the expected  time it takes for the mutation to become dominant in the population \citep{MS76,DeFi07}.
To see this, let us again resort to the branching process approach that led to \eqref{Haldane}. Namely, we approximate the expected offspring size of a mutant after $i$ days  by $Z_i$, where $(Z_i)_{i \in \mathbb{N}}$ is a discrete-time branching process  with offspring expectation $1+s$ per day,  condition on non-extinction, and obtain 
\begin{equation}\label{branching}
\begin{split}
 \mathbb E[Z_i  \mid Z_i > 0] &= \frac{\mathbb E[Z_i \, \mathbbm{1}_{Z_i>0}]}{\mathbb P(Z_i>0)} \\
 & \sim_i  \frac{\mathbb E[Z_i]}{\pi} \sim  \frac{v}{2} \, \frac{(1 + s)^i}{s}, 
\end{split}
\end{equation}
where $\mathbbm{1}$ is the indicator function, $\sim_i$ means asymptotic equivalence as $i \to \infty$ (while $\sim$ continues to refer to $N \to \infty$), and \mbox{$\mathbb P (Z_i>0) \sim_i \pi$} because extinction typically happens early; whereas for large $i$,  the process has grown  large  with high probability and then only runs a tiny extinction risk. Since $\log (1+s) \sim s$, the quantity $u$  of \eqref{u_N} is then (as $N \to \infty$) asymptotically equivalent to the solution of
\[
 \mathbb E[Z_i  \mid Z_i > 0] \sim   \varepsilon \, N
\]
for any positive constant $\varepsilon$ (so the right-hand side is a sizeable proportion of the population).

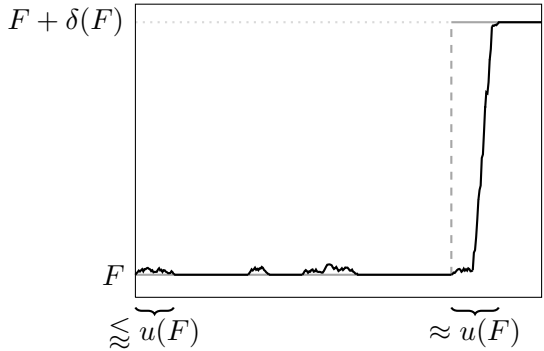
\begin{figure}\centering
\usetikzlibrary{decorations.pathreplacing}           % needed for underbraces

\newcommand{\Mut}[6]{% xoffset, points, advance, rand factor, skip, yoffset
  \foreach \y in {1,...,#5}
  { 
    \draw[black] (0,rand);
  }
  
  \draw[black,thick] (#1,#6)
  \foreach \x in {1,...,#2}
  { 
    -- ++(#3,rand*#4)
  };
}

\begin{tikzpicture}[scale=0.6]

% define style & underbraces
\tikzset{
   brace/.style={
   		thick,
     decoration={brace, mirror},
     decorate
   }
}

% --------------------------------

% limiting jumpprocess
\draw[color=gray!70, thick, dashed] (7.0, 0)   -- (7.0,5.6);
\draw[color=gray!70, thick]         (7.0, 5.6) -- (9,5.6);
\draw[color=gray!70, thick]         (0,   0)   node[left, black]{$F$}                -- (7.0, 0);
\draw[color=gray!30, thick, dotted] (0,   5.6) node[left, black]{$F + \delta(F)$}  -- (7.0,5.6) ;

% --------------------------------

% frame
\draw[-] (0, -.5) -- (9, -.5) node[right, below]{} -- (9,6) -- (0,6) -- (0,-.5);

% --------------------------------

% constants
\def\jumpwidth{0.025}
\def\jumpheight{0.05}

% --------------------------------

% underbraces
\node (startD) at (0, -.5) {};
\node (endD)   at (34 * \jumpwidth, -.5) {};
\node (startS) at (7, -.5) {};
\node (endS)   at (7 + 42 * \jumpwidth, -.5) {};

\draw [brace] (startD.south) -- node[below] {$\lessapprox u(F)$} (endD.south);
\draw [brace] (startS.south) -- node[below] {$\approx u(F)$}     (endS.south);

% --------------------------------

% first mutant try
\pgfmathsetseed{300}
\Mut{0}{34}{\jumpwidth}{\jumpheight}{50}{0}
\draw[black,thick] (0   + 34 * \jumpwidth, 0.02) -- (0   + 35 * \jumpwidth, 0) -- (2.5, 0);

% second mutant try
\pgfmathsetseed{300}
\Mut{2.5}{18}{\jumpwidth}{\jumpheight}{126}{0}
\draw[black,thick] (2.5 + 18 * \jumpwidth, 0.02) -- (2.5 + 19 * \jumpwidth, 0) -- (3.7, 0);

% third mutant try
\pgfmathsetseed{300}
\Mut{3.7}{48}{\jumpwidth}{\jumpheight}{333}{0}
\draw[black,thick] (3.7 + 48 * \jumpwidth, 0.04) -- (3.7 + 49 * \jumpwidth, 0) -- (7,   0);

% succesful mutant goes to fixation
\pgfmathsetseed{444}
\Mut{7}{19}{\jumpwidth}{\jumpheight}{0}{0}

\pgfmathsetseed{444}
\Mut{7 + 19 * \jumpwidth}{5}{\jumpwidth}{0.62}{436}{0.10}

\pgfmathsetseed{444}
\Mut{7 + 24 * \jumpwidth}{6}{\jumpwidth}{0.62}{442}{1.66}

\pgfmathsetseed{444}
\Mut{7 + 30 * \jumpwidth}{4}{\jumpwidth}{0.62}{450}{3.7}

\pgfmathsetseed{444}
\Mut{7 + 34 * \jumpwidth}{2}{\jumpwidth}{0.62}{452}{4.76}

\pgfmathsetseed{444}
\Mut{7 + 36 * \jumpwidth}{5}{\jumpwidth}{\jumpheight}{533}{5.5}

\draw[black,thick] (7 + 41*\jumpwidth,5.56) --  (7 + 42*\jumpwidth,5.6) -- (9,5.6);

% end

\end{tikzpicture}
\caption{
Schematic drawing of the relative fitness process (black) and the approximating jump process (grey). 
}
\label{loss_and_fix}
\end{figure}

All this now leads us to the dynamics of the relative fitness process.  As illustrated in Fig.~\ref{loss_and_fix}, most mutants only grow to small frequencies and are then lost again (due to the sampling step). But if it does happen that a mutation survives the initial fluctuations and gains appreciable frequency, then the dynamics turns into an asymptotically deterministic one and takes the mutation to fixation quickly, cf.~\cite{GL98}, \cite{DeFi07}, or \citet[Ch.~6.1.3]{Dur08}.  Indeed, within  time  $u(F)$, the mutation has either disappeared or gone close to fixation. Moreover, in the scaling regime \eqref{scaling} specified in the next paragraph, this time is much shorter than the mean interarrival time $1/\mu$ between successive beneficial mutations (recall that $\mu$ is the \emph{sample-wide} mutation probability).  As a consequence, there are, with high probablity, at most two types present in the population at any given time (namely, the \emph{resident} and the \emph{mutant}), and \emph{clonal interference is absent}. Therefore, in the scenario considered, survival of  drift  is equivalent to fixation. In the literature, the parameter regime $u \ll 1/\mu$ is known as the \emph{periodic selection} or \emph{sequential fixation} regime, and the resulting class of \emph{origin-fixation} models is reviewed in \cite{McCS14}. 

Next, we consider the expected per-day increase  in relative fitness, given the current value~$F$. This is
\begin{equation}\label{E_F}
\begin{split}
\EE(\Delta F \mid  F) \, & \approx \, \mu \, \pi(F) \, \delta (F) \\
& \sim \, \frac{\Gamma}{F^{2q+1}}.
\end{split}
\end{equation}
Here, the asymptotic equality is due to \eqref{delta_N}--\eqref{s_N} and \eqref{pi_N}, and the compound parameter
\begin{equation}\label{compound}
\Gamma := C \, \mu \, \varphi^2 \log \gamma
\end{equation}
is the rate of fitness increase per day at day~0 (where $r_{0j}^{} \equiv F_0^{}= 1$). Note that $\varphi/F^q$ appears squared in the asymptotic equality in \eqref{E_F} since it enters both $\pi$ and $\delta$. Note also that the additional $+1$ in the exponent of $F$ comes from the factor of $1/F$ in the length of the growth period \eqref{sigma}, and thus reflects the runtime effect. As was explained in the context of \eqref{s} and \eqref{Haldane}, this crucial difference is caused by the decrease of the selective advantage with decreasing  length of the growth period.  We will analyse this difference in some  depth in the Discussion. Let us only mention here that the effect would be absent if, instead of our Cannings model, a  discrete-generation scheme were used, as by \citet{WRL13}; or a standard Wright-Fisher model,  for which \citet{KTP09}  calculated the expected fitness increase  and the fitness trajectory for various  fitness landscapes, including the one given by \eqref{delta_N}. 
Eq. \eqref{E_F} now leads us to  define  a new time variable $\tau$ related to $i$ of \eqref{t_gamma} via
\begin{equation}\label{tau_Gamma}
i = \Big \lfloor \frac{\tau}{\Gamma} \Big \rfloor
\end{equation}
with $\Gamma$ of \eqref{compound}, which means that one unit of time $\tau$ corresponds to $\Gamma$ days. With this rescaling of time, Eq. \eqref{E_F}
%we have
%\[
%F^{N}_{\lfloor  \tau / \Gamma_N \rfloor} \to f(\tau) \text{ as } N \to \infty, 
%\]
%where $f$ satisfies the   initial value problem 
corresponds to the differential equation 
\begin{equation}\label{ODE}
\frac{\D{}}{\D{\tau}} f(\tau) = \frac{1}{f^{2q+1}(\tau)}, \quad f(0)=1,
\end{equation}
with solution
\begin{equation}\label{LLN}
  f(\tau) = \big ( 1+2\, (1+q) \, \tau \big )^{\frac{1}{2(1+q)}}.
\end{equation}
This is the desired power law for the fitness trajectory.
Note that \eqref{ODE} is just a scaling limit of \eqref{E_F}, where the expectation was  omitted due to a dynamical law of large numbers, as will be explained next.
\paragraph{Scaling regime and law of large numbers.} We now think of $\mu = \mu_N$ and $\varphi= \varphi_N$ as being indexed with population size because the law of large numbers requires to consider a sequence of processes indexed with $N$. Thus, other   quantities now also depend on $N$ (so $\delta=\delta_N^{}, s=s^{}_N$, $\pi=\pi^{}_N$, $\Gamma=\Gamma_N$ etc.), and so does the  \emph{relative fitness process}  $(F_i)_{i \geqslant 0} = (F^N_i)_{i \geqslant 0}^{}$ with $F_i$ of \eqref{rel_fitness}. More precisely, we will take a \emph{weak mutation--moderate selection limit}, which requires that $\mu_N$ and $\varphi_N$ become small in some controlled way as $N$ goes to infinity. Specifically, \citet{GKWY16} assume
\begin{equation}\label{scaling}
\begin{split}
& \mu^{}_N \sim \frac{1}{N^{a}}, \; \varphi^{}_N \sim \frac{1}{N^{b}} \quad \text{as} \quad N \to \infty, \\
& 0 < b < \frac{1}{2}, \;  a > 3\,b.
\end{split}
\end{equation}
Due to the assumption $a> 3\,b$, $\mu_N$ is of much lower order than $\varphi_N$.  This is  used by  \citet{GKWY16} to prove that, as $N\to \infty$,   with high probability  no more than two fitness classes are simultaneously present in the population over a long time span. Note that $\mu_N$ is the per-day \emph{mutation probability per population} (but see the discussion at the end of the  paragraph on interday dynamics at the beginning of this section).
  
Furthermore, the scaling of $\varphi_N$ implies that selection is stronger than genetic drift as soon as the mutant has reached an appreciable frequency. The method of proof applied by  \citet{GKWY16} requires the assumptions \eqref{scaling} in order to guarantee a coupling between the new mutant's offspring and two nearly critical Galton-Watson processes between which the mutant offspring's size is `sandwiched' for sufficiently many days. Specifically, under the assumption $0 < b < \frac{1}{2}$,  the coupling  works  until the mutant offspring in our Cannings model has reached a small    (but strictly positive) proportion of the population, or has disappeared. A careful inspection of the arguments shows that, under the weaker condition $0 < b < \frac{2}{3}$, this coupling works at least until the mutant offspring has (either disappeared or) reached  size $N^b$, from which it then goes to fixation by a law of large numbers argument. This makes the limit result of \citet{GKWY16} valid for $0 < b < \frac{2}{3}$; we conjecture that it even holds for $0 < b < 1$.

In the case where selection is much stronger than mutation, the classical models of population genetics, such as the Wright-Fisher or Moran model, display the well-known dynamics of sequential fixation. Two distinct  scenarios can happen (see the review by \citet{McCS14}, or \citet[Ch.~2 and Fig.~2.7]{GL00}): either a fast loss of a new beneficial mutation, or its fixation. Qualitatively, our Cannings model displays a  similar behaviour.
Furthermore, as already indicated, with the chosen scaling the population turns out to be homogeneous on generic days~$i$ as $N\to \infty$. This has the following  practical consequences for the relative fitness process $(F_i^N)_{i \geqslant 0}^{}$.
First, on a time scale with a unit of $1/(\mu_N \, \varphi_N)$ days, $(F^N_i)_{i \geqslant 0}^{}$ turns into a jump process as $N \to \infty$, cf. Fig.~\ref{loss_and_fix}. 
 Second,  on the (generic) days $i$ at which the populations are nearly homogeneous,  the  subtle systematic difference between \eqref{emp_rel_fitness} and \eqref{rel_fitness}, as described in Footnote~\ref{fn:compare:emp_rel_fit:rel_fitness}, will disappear.
The precise formulation of the limit law \citep{GKWY16} reads as

\smallskip
\noindent
{\bf Theorem}
{\it For $N \to \infty$ and under the  scaling \eqref{scaling}, the sequence of processes $\big (F^N_{\lfloor \tau /\Gamma_N \rfloor} \big )_{\tau \geqslant 0}$ converges,
%for $N \to \infty$, 
in distribution and locally uniformly, to the deterministic function $\big ( f(\tau) \big )_{\tau \geqslant 0}$ in \eqref{LLN}.}
%\end{theorem}

\smallskip
\noindent

The theorem was proved  along the heuristics outlined above\footnote{Note that \cite{GKWY16} partly work with dimensioned variables, which is why the notation and the result look somewhat different.} 
with the help of advanced tools from  probability theory. 
%It is a law of large numbers in the sense that, for large $N$ and under the scaling assumption \eqref{scaling}, fitness is the sum of a large number of  small per-day  increments accumulated over many days, and may  be approximated by its expectation. It is this kind of reasoning which allows to go  from \eqref{E_F} to \eqref{ODE} (and thus allows us to get rid of the expectation operator). 
It is a law of large numbers reasoning, which allows to go from \eqref{E_F} to \eqref{ODE} (and thus to `sweep the expectation under the carpet'), in the following sense: For large $N$ and under the scaling assumption \eqref{scaling}, fitness is the sum of a large number of small per-day increments accumulated over many days, and may be approximated by its expectation. 

Since time has been rescaled via \eqref{tau_Gamma}, Eq.~\eqref{LLN} has $q$ as its single parameter. Note that  $1/(2\,(1+q))<1$ (leading to a concave $f$)  whenever $q \geqslant 0$; in particular, \emph{the fitness curve is concave even for} $q=0$, \emph{that is, in the absence  of epistasis}. 
In contrast, the fitness trajectory obtained by \cite{KTP09}  for the Wright-Fisher model under $q = 0$ is linear. The difference is due to the runtime effect, which is present in our Cannings model even for $q=0$ because of the parametrisation of the intraday dynamics with the individual reproduction rate $r$: If the population as a whole already reproduces faster, then the end of the growth phase is reached sooner and thus leaves less time for a mutant to play out its advantage $\delta(r) = \varphi/r^0 = \varphi$ of \eqref{r_increase};  see also the discussion in Sec.~\ref{sec:discussion}. The Wright-Fisher model of \cite{KTP09}  does not display the runtime effect because it does not contain the individual (intraday) reproduction rate as a parameter.

The second parameter, namely  $\Gamma_N$, reappears when  $\tau$ is translated back into days; that is,  $F^N_i \approx f(\Gamma_N \, i)$. Note that  $R_0$, as used in the first nondimensionalisation step \eqref{t}, is not an additional parameter because it is already absorbed in $\varphi^2_N$. 

\begin{figure*}[!ht]\centering
\resizebox{.63\textwidth}{!}{
\input{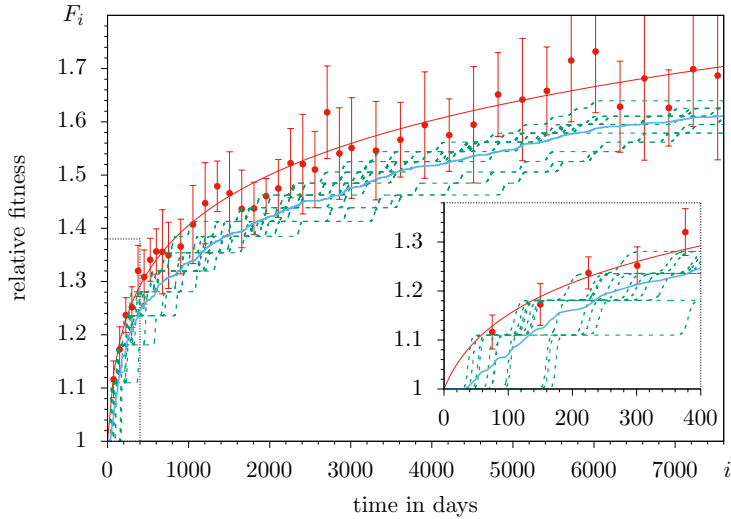}
}
\caption{Least-squares fit of the curve \eqref{LLN} to the data in \cite{WRL13DATA}, and stochastic simulations of finite populations with deterministic beneficial effects. Red bullets: mean empirical relative fitness  (averaged over all 12  populations) with error bars as in Fig.~\ref{fig:WRL_data}; solid red line: $F^{}_i \approx f(\widehat\Gamma \, i)$ with parameter values $\widehat q=4.2$  and $\widehat \Gamma=3.2 \cdot 10^{-3}$; green lines: 12 individual trajectories $F_i$ obtained via Cannings simulations with $N=5 \cdot 10^6, \gamma=100, \widehat \varphi = 0.11$, and $\widehat \mu=0.057$; light blue line: average over the 12 simulations; inset: zoom on the early phase.}
\label{cannings_det_init_0p11_uncorrected}
\end{figure*}

\paragraph{A first reality check.}  The  limit law \eqref{LLN} is identical with  the power law \eqref{powerlaw} of \cite{WRL13} up to a transformation of the parameters that relies on relevant details in the modelling (see also the discussion in Sec.~\ref{sec:discussion}). We have $q=g-1$, so $\widehat g = 5.2$ of Sec.~\ref{sec:LTEEWRL} translates into $\widehat q=4.2$.\footnote{Recall that we denote parameter estimates by a hat to distinguish them from the corresponding theoretical quantities.}  Furthermore, $\Gamma = (\beta \, \log_2 \gamma) /(2\, (1+ q))$  due to \eqref{powerlaw} and \eqref{LLN} together with the fact that $k = (\tau \log_2 \gamma)/\Gamma$ by  \eqref{t_gamma} and \eqref{tau_Gamma}; given %$\widehat \beta=0.00515$
$\widehat \beta=5.1 \cdot 10^{-3}$, this results in $\widehat \Gamma = 3.2 \cdot 10^{-3}$ (here and in what follows, we again suppress the index $N$, since  we will work with fixed, finite $N$ from now on). The resulting fit is reproduced in Fig.~\ref{cannings_det_init_0p11_uncorrected} (red solid line). 
In line with \citet[Fig.~2]{WRL13}, we average over all 12 populations, 
at this point neglecting a certain variability of the parameters between the populations, see their Table~S4.  For comparison, we have also  included in Fig.~\ref{fig:WRL_data} the fit without epistasis, that is, for $q=0$; as well as the linear one, which applies in the absence of both epistasis and runtime effect. In the notation corresponding to \eqref{powerlaw},   these are  $\widetilde f(k)=\sqrt{1+\widehat{\beta}_{\mathrm{sqr}} \, k}$ and  $\widetilde f(k)=1+ \widehat{\beta}_{\mathrm{lin}} \, k$ with suitable constants $\widehat{\beta}_{\mathrm{sqr}}$ and $\widehat{\beta}_{\mathrm{lin}}$.

In the light of \eqref{compound}, of the given value $\widehat \Gamma$, and of the fact that $C \, \log \gamma  \approx 4.7$, the values of $\widehat \mu$ and $\widehat \varphi$ cannot both be very small. We therefore now check the limit law against realistic parameter values.

We start by decomposing the compound parameter $\Gamma$. Recall from \eqref{r_increase} that the \emph{fitness increment due to the first beneficial mutation} is
\begin{equation}
\label{delta1}
	\varphi = \delta(F_0^{}) = \delta(1).
\end{equation}	
This was estimated as $0.1$ by \citet{Lenski91}, see also \citet{GL98}, and \citet{WRL13}. For reasons to be explained in Sec.~\ref{sec:det}, however, we work with the somewhat larger value $\widehat \varphi = 0.11$. The mutation probability may then be obtained from \eqref{compound} as
\begin{equation}\label{est_mutrate}
\widehat \mu=\frac{\widehat\Gamma}{C \, \widehat \varphi^{\,2} \log \gamma} = 
0.057.
\end{equation}
Stochastic simulations of the  GKWY model, performed with Algorithm~\ref{alg:cannings} described in Appendix~C and using the above parameters\footnote{The table in Appendix~C contains the precise values used in the simulations, whereas numbers are rounded to two decimals throughout the text.} together with {$N= 5 \cdot 10^6$}, are also shown in Fig.~\ref{cannings_det_init_0p11_uncorrected}. Their mean (over 12  runs) recovers the basic shape of the fitness curve, but systematically underestimates both the limit law and the data.  A natural explanation for this is clonal interference, which is absent in the limit under the scaling \eqref{scaling}, but leads to loss of mutations for finite $N$. This will be taken into account in Sec.~\ref{sec:ci}. But let us note here that the fluctuations in the data are rather larger than those of the simulations; this may well go along with a variability of the parameters between the 12 replicates of the LTEE, which is present in the data, but not in our simulations.

\section{Including clonal interference} 
\label{sec:ci}
As discussed in Sec.~\ref{sec:GKWYLLN}, the scaling regime in the GKWY model was such that, with high probability, no new beneficial mutation arrived while the previous one was on its way either to extinction or fixation. As indicated by the simulation results  in Fig.~4, also clonal interference should be taken into account. Briefly stated, clonal interference refers to the situation where a second contending mutation  appears while the previous one is still on its way to fixation (recall also Footnote~\ref{foot_CI}). It is crucial to keep in mind that, unlike the case without clonal interference considered in Sec.~\ref{sec:GKWYLLN}, survival of  drift may then no longer be identified with fixation;  rather, there may be an additional loss of contending mutations   due to clonal interference. In particular, the quantity $\pi$ of \eqref{pi_N} must now be addressed as the \emph{probability to survive drift} rather than the fixation probability.

A full analytic treatment of clonal interference is beyond the scope of this paper; in particular, we will not prove a law of large numbers here. Rather, we refine and adapt the heuristics of \cite{GL98}, see also \cite{WRL13}. We will first consider the deterministic effects as assumed in the GKWY model in Sec.~\ref{sec:det} and then proceed to random effects from a very general class of probability distributions in Sec.~\ref{sec:stoch}.

\subsection{Deterministic beneficial effects}
\label{sec:det}
The heuristics of \citet{GL98} was originally formulated for fitness effects that follow an exponential distribution;
if applied to the degenerate case of derministic   effects, it leads to certain artifacts. We  will therefore sketch and apply a {\em thinning heuristics} as a counterpart to the Gerrish-Lenski heuristics. Consider the situation that a second mutation surviving drift appears within the time window $u(F)$ of \eqref{u_N} after the appearance of a first  mutation (this is more or less while the first mutation has not become dominant yet). Then, with high probability, the second mutation occurs in an individual of relative fitness $F$ (rather than in an individual of relative fitness $F+\delta(F)$), and therefore belongs to the same fitness class as the first mutant and its offspring. Thus, as far as fitness is concerned, the two mutants (and their offspring) can be considered equivalent. In our heuristics, the occurrence of a second (and also a third, fourth, $\ldots$) mutation within the given time window neither speeds up nor decelerates the (order of magnitude of) the time until the new fitness class is established in the population. So $u(F)$ plays the role of a \emph{dead time}, in the sense that the fitness increments carried by contending mutations arriving within this period are lost. We now determine the probability that a given increment is \emph{not} lost 
by comparing the intensities of two point processes. The first is the process of contending mutations arriving at rate (or intensity) $I_1(F) := \mu \, \pi(F)$. The second is the process of contending mutations arriving outside the dead time of the preceding one. This is a renewal process, where the next point appears after a waiting time of $u(F)$ plus an $\Exp(I_1(F))$-distributed random variable\footnote{Note that both processes are in \emph{continuous time} and approximate what happens in the original discrete-time model. In this sense, $\mu$ is to be understood as a mutation \emph{rate} here.}. The intensity of this process is then the inverse of the expected waiting time, namely 
\[
I_2 (F) := \frac{1}{u(F)+1/I_1(F)}.
\]
By a simple argument from renewal theory (see e.g. \cite[Ch.~3.4, in particular Ex.~4.3]{Dur05}), the fraction of contending mutations that are not lost due to clonal interference  at fitness level $F$ is thus approximately given by the {\em retainment factor}
\begin{equation}\label{survive_ci_det}
\frac{I_2(F)}{I_1(F)} = \frac{1}{1 + I_1(F) \, u(F)}=:\vartheta(F).     
\end{equation}
Under this approximation, the expected per-day increase of the relative fitness, given its current value $F$, turns into
\begin{equation}\label{withtheta}
\EE(\Delta F  \mid  F) \,
 \approx \, \mu \, \pi(F) \, \delta (F)\,\vartheta(F).
\end{equation}
We recall from \eqref{delta_N}--\eqref{s_N} that
\[
%\begin{equation}\label{sF}
s(F) =  \frac{\varphi\log\gamma}{F^{q+1}}.
%\end{equation}
\]
Hence \eqref{u_N} becomes 
\begin{equation}\label{ufull}
u(F)\sim \frac {\log \big ((N \, \varphi \log \gamma)/F^{q+1} \big )}{s(F)};
\end{equation}
and, due to \eqref{pi_N}, the effect of $F$ cancels out in the leading term in the product of $\pi(F)$ and $u(F)$ in \eqref{survive_ci_det}.  Put differently,  the dead time and the expected interarrival time of contending mutations increase with $F$ in the same way, up to logarithmic corrections.
 
In order to dispose of the remainig dependence of $F$, namely the one in  the numerator of  \eqref{ufull}, we replace the factor  $(\log \gamma) / F^{q + 1}$  by 1. This somewhat crude-looking approximation seems justified because the term appears under the logarithm in \eqref{ufull}, and $\log(\log(\gamma) / \widetilde{F}^{\widehat q + 1})$  is between $-1$ and $+1$ for $\gamma=100$, our estimate $\widehat q=4.2$,  and $\widetilde F$ between $1.1$ and $1.6$ (recall from Fig.~\ref{cannings_det_init_0p11_uncorrected} that $\widetilde F$ is between 1 and 1.7). On the other hand, the estimated value of $\log (N \,  \varphi)$ is $ \log (N \,  \widehat \varphi) = 13.22$ for $N = 5 \cdot 10^6$ and $\widehat \varphi=0.11$. 
With this approximation, \eqref{ufull} turns into
\begin{equation}\label{ulight}
u(F)\approx \frac {\log (N \, \varphi)}{s(F)}
\end{equation}
and \eqref{survive_ci_det} becomes
\begin{equation}\label{constant_theta}
\vartheta(F) \equiv \vartheta =  \frac{1}{1+C \, \mu  \, \log(N \, \varphi)}.
\end{equation}
Moreover, \eqref{withtheta} becomes
\begin{equation}\label{Exp_incr_F_i}
\EE(\Delta F  \mid  F) \,
 \approx   \frac{\Gamma}{F^{2q+1}},
\end{equation}
where now
\begin{equation}\label{compound2}
\Gamma =  \frac{C \,\mu \, \varphi^2 \log \gamma}{1+ C \, \mu \log (N \, \varphi)} ,
\end{equation}
that is, the factor $\mu$ in \eqref{compound} is replaced by  $\mu/(1+C \, \mu \log (N \, \varphi))$. Now, taking the expectation over $F$ in \eqref{Exp_incr_F_i} yields
\[
\EE(\Delta F )  \, \approx  \Gamma \,
\EE \Big ( \frac{1}{F^{2q+1}} \Big ) 
%\gtrapprox  \frac{\Gamma}{\big (\EE(F)\big )^{2q+1}}
. 
\]
%Here, the second step is due to Jensen's inequality.\footnote{Note that $1/x^p$ is a convex function of $x$ for any $p \geqslant 1$.} 
Assuming a suitable concentration of the random variables in question around their expectations (which in theory would be justified by a dynamical law of large numbers result such as the one discussed in  Sec.~\ref{sec:GKWYLLN}, and in practice is a crude way of moment closure also implied by \citet{WRL13} and \cite{KTP09}), we interchange the expectation with the nonlinearity and  arrive at
%The interchange of the expectation with this function reflects a crude way of moment closure. We therefore arrive at 
the approximation
\[
F_{\lfloor   \tau / \Gamma \rfloor} \approx \EE \big ( F_{\lfloor  \tau / \Gamma \rfloor}  \big ) \approx f(\tau)  \; \text{for large } N
\]
with $f$ as in \eqref{LLN}. We may, therefore,  approximate (as in Fig. \ref{cannings_det_init_0p11_uncorrected}) the data by the function $f$, with the same  values $\widehat q$ and $\widehat \Gamma$ as before. The compound parameter $\Gamma$, however, has an internal structure different from the  previous one (compare \eqref{compound2} with \eqref{compound}). 

\begin{figure}
\resizebox{\columnwidth}{!}{\input{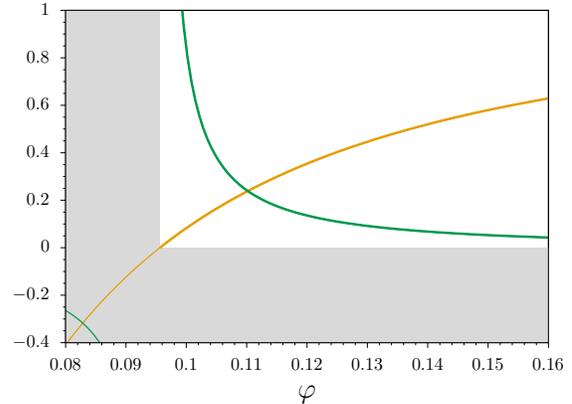}}
\caption{\label{fig:pole} $\mu$ (green) and $\vartheta$ (orange) as functions of $\varphi$ according to \eqref{find_mu} and \eqref{constant_theta}; the former has a pole at $\approx 0.096$. Values of $\varphi \lessapprox 0.96$ are forbidden  because there  $\mu$ (and $\vartheta$) would become negative.}
\end{figure}

Solving  \eqref{compound2} for $\mu$ now yields the  mutation rate 
\begin{equation}\label{find_mu}
    \mu = \frac{\Gamma}{C  \, \big (  \varphi^2 \log \gamma -  \Gamma \, \log (N \,  \varphi)\big )} .
\end{equation}
However, the denominator has a pole at \mbox{$\varphi \approx 0.096$} (and is negative for smaller values of $\varphi$), see Fig.~\ref{fig:pole}.
The existence of the pole, and the resulting explosion of $\mu$ in its neighbourhood, have the following meaning. 
%reflects the fact that, for small $\varphi$, there is an upper limit to the rate of evolution under a sequential fixation assumption including the thinning heuristics. This is because small $s$ implies large $u$, and the sum of the dead times defines the minimal spacing of the mutations that can go to fixation. This minimal spacing is achieved in the limit $\mu \to \infty$, which entails $\vartheta \to 0$. 
According to \eqref{ulight}, the window length $u(F)$ depends   on $N, \varphi$, and $s (F)$. Each window goes along with an increment of $F$ by $\delta(F)$, and a spacing on the time axis until the next window begins. For smaller $\varphi$,  the increments $\delta(F)$ become smaller and the windows get wider, which inevitably means that the gaps between the windows have to be shorter (in order to obtain the  observed total increase of $F$ of $\approx 0.7$ within the given time). Shorter gaps between the windows, however, mean larger mutation rates and, in the limit of vanishing gaps, even an infinite mutation rate (and a vanishing retainment factor), which is, of course, not realistic.
For an asymptotic analysis as $N\to \infty$, this  suggests one has to assume that $\vartheta$ is bounded away from 0. For substantially higher mutation probabilities, the heuristics would break down \citep{FND08} and a different asymptotic regime would apply \citep{DeFi07,DM11}. 

\begin{figure*}[!ht]\centering
\resizebox{.63\textwidth}{!}{\input{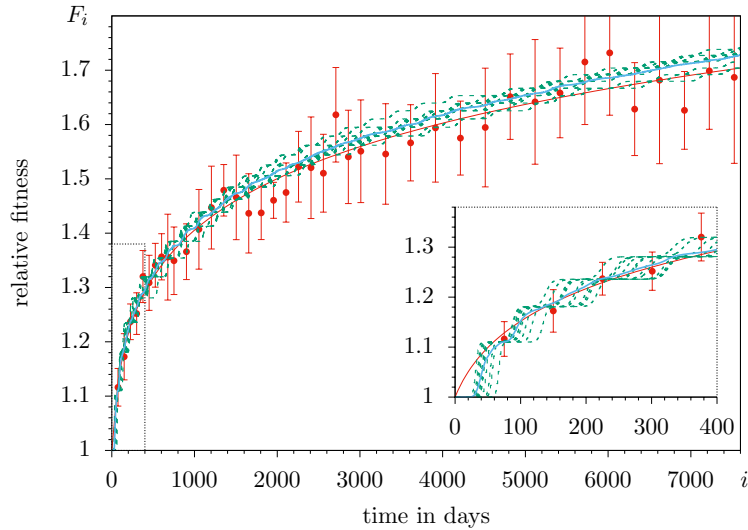}}
\caption{Cannings simulation as in Fig.~\ref{cannings_det_init_0p11_uncorrected}, but with mutation probability $\widehat \mu=0.24$. %0.242
}
\label{cannings_det_init_0p11}
\end{figure*}

Our choice of $\widehat \varphi \approx 0.11$ for the simulations in Section~\ref{sec:GKWYLLN} was intended to avoid the numerical instabilities close to the pole of \eqref{find_mu}.
For this value, Eq.~\eqref{find_mu}  gives $\widehat \mu=0.24$ and, via \eqref{constant_theta}, a retainment factor of $\widehat \vartheta=0.24$. %0.236
As was to be expected, this now gives a better agreement between the simulated mean fitness and the approximating power law (and hence with the data), see Fig.~\ref{cannings_det_init_0p11}. 

\begin{figure*}[!ht]\centering
\resizebox{.63\textwidth}{!}{\input{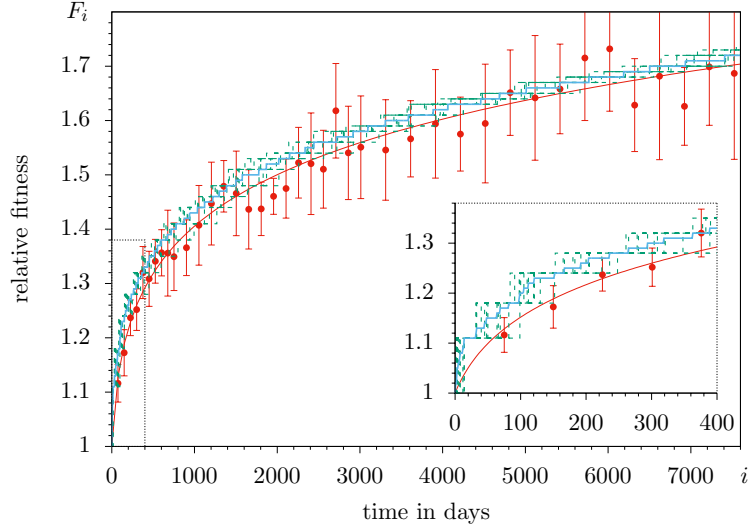}}
\caption{Simulation using  heuristics for deterministic increments. Parameters as in Fig.~\ref{cannings_det_init_0p11}. 
Mean number of clonal interference events: 84; mean number of established beneficial mutations: 27.
}
\label{approx_det_init_0p11}
\end{figure*}

Recall that the parameters $\widehat \mu$ and $\widehat \varphi$ have been obtained by fitting a first order (ODE) approximation (of the above described {\em thinning heuristics}) to the empirical data, taking into account some information on the effect of the first successful beneficial mutation.  As a consistency check, it is interesting to also simulate the thinning
heuristics with these parameters (see Algorithm~\ref{alg:approximation}
in Appendix~C) and compare the result with the simulations based on the
Cannings model. As shown in Fig.~\ref{approx_det_init_0p11}, the fit of
the mean is better for the Cannings simulations than for the simulation
of the  heuristics in the early phase of the LTEE, and vice versa in the late phase. Note that the simulations of the heuristics yields smaller fluctuations than that of the Cannings model; this goes along with the fact that the model based on the heuristics contains fewer random elements than the Cannings model.

 With the parameter values $\widehat \varphi = 0.11$ and $\widehat \mu  = 0.24$, the number of fixed beneficial mutations in the simulation in Fig.~\ref{approx_det_init_0p11}, averaged over the 12 runs, is 27; this is to be compared with the estimate of 60--110 fixed mutations observed in 50000 generations by \citet{Tenaillon16}, and of 100 fixed mutations observed in 60000 generations by \citet{Good17}, which both include neutral mutations and mildly deleterious hitchhiking `passenger' mutations. We will come back to this in the discussion.

\subsection{Random beneficial effects}
\label{sec:stoch}
Let us now turn to random beneficial effects. To this end, we scale the fitness increments with a positive random variable $X$ with density $\density$ and expectation $\EE(X)=1$. We assume throughout that  $\EE(X^2) < \infty$ to ensure that all  quantities required in what follows are well-defined.

Taking into account the dependence on $X$, the quantities in \eqref{delta_N}--\eqref{s_N}, \eqref{pi_N} and \eqref{u_N} 
turn into
\begin{subequations}\label{X}
\begin{linenomath}\postdisplaypenalty=0
\begin{align}
\delta (F,X) & = X \, \frac{\varphi}{F^q}, \label{X:delta}\\
\sigma(F) & = \, \frac{\log \gamma}{F} \text{ (as before)}, \label{X:sigma}\\
s(F,X) \, & = \, \delta(F,X) \, \sigma(F), \label{X:s}\\
\pi(F,X) \, & \approx \, C \, s(F,X), \label{X:pi}\\
\begin{split}
u (F,X) \, & = \, \frac{\log (N\,s(F,X))}{s(F,X)} \\ &\approx \frac{\log (N\, \varphi\, X)}{s(F,X)}. \label{X:u}
\end{split}
\end{align}
\end{linenomath}
\end{subequations}

In \eqref{X:u} we apply the same reasoning that led to the approximation~\eqref{ulight} for $u$.
Note  that large $X$ implies large $s$ and hence small $u$ and vice versa.
%; and second that \eqref{X:pi} is an approximation,  whereas in  \eqref{pi_N} we have asymptotic equivalence. 
The following Poisson picture will be central to our heuristics: %
% A beneficial mutation with scaled effect $x$ that arrives at time $\tau$ corresponds to a point $(\tau, x) \in \mathbb R_+\times \mathbb R_+$. The Poisson process of beneficial mutations then has intensity $\mu \D{\tau} h(x) \D{x}$, and in  fitness background $\approx F$, the Poisson process of {\em contending mutations}, i.e. those beneficial  mutations that survive drift (but not necessarily go to fixation) has  intensity $\mu \D{\tau} h(x) \pi(F,x) \D{x}$.
The process of \emph{beneficial mutations} with scaled effect $x$ that arrives at time $\tau$ has intensity $\mu \D{\tau} h(x) \D{x}$ with points $(\tau, x) \in \mathbb R_+\times \mathbb R_+$. %
And in fitness background $\approx F$, we denote by $\Pi$ the Poisson process of \emph{contending mutations}, i.e. those beneficial mutations that survive drift (but not necessarily go to fixation), which has intensity $\mu \D{\tau} h(x) \, \pi(F,x) \D{x}$ on $\mathbb R_+\times \mathbb R_+$. %
\\
We now develop a refined version of the {\em Gerrish-Lenski heuristics for clonal interference} 
%, extended to a general  distribution of beneficial effects, 
and adapt it to the context of our model.
If, in the fitness background $\approx F$, two contending mutations $(\tau, x)$ and $(\tau', x')$ appear at $\tau < \tau' < \tau + u(F,x)$, then the first one outcompetes (`kills') the second one if $x' \leqslant x$, and the second one kills the first one if $x' > x$. Thus, neglecting interactions of higher order, given that a contending mutation arrives at $(\tau, x)$ in the fitness background $\approx F$, the probability that it does not encounter a killer in its  past is %
%
%\begin{linenomath}
%\begin{align}
%&	\exp \Big ( -  \int_x^\infty \!\!\!  \mu \, \pi (F,y) u(F,y) \density(y) \dd y \Big ) \nonumber \\
%\end{align}
%\end{linenomath}
\begin{linenomath}
\begin{equation}
\begin{split}
\back{\chi}(& F, x) := \\
& \exp \Big ( -  \int_x^\infty \!\!\!  \mu \, \pi (F,y)\,  u(F,y) \, \density(y) \D{y} \Big ), \label{past}
\end{split}
\end{equation}
\end{linenomath}
whereas the probability that it does not encounter a killer in its future is 
\begin{equation}
\begin{split}
& \forw{\chi}(F, x) := \\
&	\exp \Big ( - u (F, x) \int_x^\infty \!\!\!  \mu \, \pi (F,x') \, \density(x') \D{x'} \Big )
\end{split}
\label{future}
\end{equation}
(note that only the term corresponding to $\forw{\chi}$ is considered by \citet{GL98}).
Using \eqref{X}, $\back{\chi}(F, x)$ is approximated by
\begin{linenomath}\postdisplaypenalty=0
\begin{align}
\back{\psi}(x) & := \notag \\
\exp & \Big ( -\mu\, C   \int_x^\infty  \log (N \, \varphi \, y) \density(y) \D{y}  \Big ), \label{psipast}
\intertext{whereas $\forw{\chi}(F, x)$ is approximated by}
\begin{split}
\forw{\psi}(x) & := \\
 \exp & \Big ( -\mu  \, \frac{C \log (N \, \varphi \, x)}{x} \int_x^\infty x' \,  \density(x') \D{x'}  \Big ).
\end{split}
\label{psifuture}
\end{align}
\end{linenomath}
In analogy with the approximation of the retainment factor \eqref{survive_ci_det} that uses \eqref{ulight}, neither $\back{\psi}$ nor $\forw{\psi}$ depend on $F$. %
Thus, setting $\bafo{\chi} := \back{\chi} \, \forw{\chi}$ and analogously $\bafo{\psi}:= \back{\psi} \, \forw{\psi}$, we obtain, as an analogue of~\eqref{Exp_incr_F_i}, the expected (per-day) increase of $F$, given the current value of $F$, as
\begin{linenomath}\postdisplaypenalty=0
\begin{align} 
& \EE  (\Delta F \mid  F) \notag \\
 &\approx \,  \mu \int_0^\infty \!\!\! \delta (F,x)  \, \pi(F,x) \,
\bafo{\chi}(F, x) \, h(x) \D{x}  \notag \\
& \approx \,   \frac{C \, \mu\, \varphi^2 \log \gamma }{F^{2q+1}} 
\int_0^\infty x^2 \, \bafo{\psi}(x) \,  \density(x) \D{x}   \label{exp_incr_random} \\
& = \, \frac{\Gamma}{F^{2q+1}}, \notag
\end{align}
\end{linenomath}
where
\begin{equation}
  \Gamma := C \, \mu \, \varphi^2 \log (\gamma) \, I(\mu, \varphi) 
  \label{compound_stoch}
\end{equation}
and $I(\mu, \varphi) := \EE \big (\bafo{\psi}(X) \, X^2 \big)$ is the integral in~\eqref{exp_incr_random}. %
Similarly as in Sec.~\ref{sec:det}, the assumption of a suitable concentration of the random variable $\Delta F$ around its conditional expectation allows us to take \eqref{exp_incr_random} into
%and thus allowing to interchange the expectation (and, in a suitable paramenter regime, presumably also a law of large numbers reasoning as in Sec.~\ref{sec:GKWYLLN}) leads to
%
\[
F_{\lfloor   \tau / \Gamma \rfloor} \approx \EE \big (F_{\lfloor   \tau / \Gamma \rfloor}  \big ) \approx f(\tau) 
\]
with $f$ as in \eqref{LLN}. %
As in Section~\ref{sec:det}, we will refer to this approximation step as `moment closure'. 
The analysis so far allows to conclude that, as long as the above-described approximation 
may be relied on, 
the \emph{power law of the mean fitness curve} observed by \citet{WRL13} \emph{is obtained under any suitable distribution of fitness effects}; in particular, the \emph{epistasis parameter} $q$ is \emph{not affected by the  distribution} of $X$. %

\paragraph{More general forms of epistasis.}
If \eqref{X:delta} is replaced by the more general condition
\begin{equation}
\delta (F,X)  = X \, \varphi\, \eta(F)\label{X:delta_general}
\end{equation}
for some (continuously differentiable, decreasing) function $\eta$ with $\eta(1)=1$ such that again the approximation \eqref{ulight} makes sense, then all the arguments in the previous paragraph go through, and we obtain 
\[ \EE  (\Delta F \mid  F)  = \, \Gamma\, \frac{ \big (\eta(F) \big )^2}{F}\]
with $\Gamma$ as in \eqref{compound_stoch}. Again, under a suitable concentration assumption, $F_{\lfloor   \tau / \Gamma \rfloor}$  is approximated by $f(\tau)$, where now $f$ solves the initial value problem
\begin{equation}\label{GODE}
\frac{\D{}}{\D{\tau}} f(\tau) = \frac{\big( \eta \big (f(\tau) \big )\big )^2}{f(\tau)}, \quad f(0)=1.
\end{equation}
For the scaling regime \eqref{scaling}, which excludes clonal interference in the limit $N\to \infty$, a corresponding dynamical law of large numbers leading to the limiting ordinary differential equation \eqref{GODE}   was proved in \citet[Cor. 2.15]{GKWY16}  for  $F_{\lfloor   \tau / \Gamma \rfloor}$ with $\Gamma$ as in \eqref{compound}.

\paragraph{Estimation of parameters.} Our next goal is to estimate the parameters $\mu$ and $\varphi$ that will then be used in simulations to check consistency as in Section~\ref{sec:det}, but now for stochastic increments. Again one starts with the composite parameter~$\Gamma$, which can be estimated from the empirical data in the same way as described at the end of Sec.~\ref{sec:GKWYLLN}. %
In Appendix~A, we derive  an approximation for  $\mathfrak d_1$, the \emph{expected effect of the first among the contending mutations} (in fitness background $F=1$) \emph{that is not killed}.\footnote{Note that in our counterpart of this heuristic for deterministic beneficial effects, $\mathfrak{d}_1$ would coincide with $\varphi$, since the first contending mutation is never killed.} 
In order to estimate $\mu$ and $\varphi$ %
from  $\widehat \Gamma$ and $\widehat{\mathfrak{d}}_1$,  the observed mean fitness increment of the first fixed beneficial mutation (in analogy with~\eqref{est_mutrate}), we combine  \eqref{X:delta1}  with \eqref{compound_stoch}  to obtain the system of equations
\begin{subequations}\label{system_est}
\begin{linenomath}\postdisplaypenalty=0
\begin{align}
\mathfrak{d}_1 & \, \approx \, \varphi \, \zeta_\ell^{}(\mu,\varphi) \label{mu_hat}, \\
\mathfrak{d}_1 & \, \approx \, \sqrt{\frac{\Gamma}{C \log \gamma}} \frac{\zeta_\ell^{}(\mu,\varphi)}{\sqrt{\mu \, I(\mu,\varphi)}}, \label{varphi_hat}
% \frac{\mu \, I(\mu,\varphi )}{(\zeta_\ell^{}(\mu, \varphi))^2} &= \frac{\Gamma}{C \, \mathfrak{d}_1^2 \log %\gamma} \label{mu_hat}, \\
%\varphi &= \frac{\mathfrak{d}_1}{\zeta_\ell^{}(\mu,\varphi)}, 
%\label{varphi_hat}
\end{align}
\end{linenomath}
\end{subequations}
where 
$\zeta_\ell(\mu,\varphi)$ is  an approximation of the expectation of the first \emph{scaled} beneficial effect that goes to fixation. The subscript $1 \leqslant \ell < \infty$ indicates the maximum number of contending mutations taken into account before the first fixation; the approximation becomes more precise with increasing $\ell$.
%$\mu$, as the solution of \eqref{mu_hat}, determines $\varphi$ via \eqref{varphi_hat}. %
We will work with $\ell = 3$ since more than three contenders turn out to be rarely present at the same time (see the polymorphism statistics in the next paragraph).  Plugging  the value $\widehat \Gamma=3.2 \cdot 10^{-3}$ (from Fig. \ref{cannings_det_init_0p11_uncorrected}) into \eqref{mu_hat}, we will solve \eqref{mu_hat} and \eqref{varphi_hat} for the parameter estimates $\widehat \mu$ and $\widehat \varphi$   for exponentially distributed $X$ in the remainder of this section. %
Let us anticipate that, as in the case of deterministic increments, one has to cope with the fact that such solutions do not exist for all values of $\widehat{\mathfrak{d}}_1$.

\paragraph{Exponentially distributed beneficial effects.} %
For definiteness, we now turn to random beneficial effects where $X$ follows $\Exp(1)$, the exponential distribution with parameter~1. %
This was the canonical choice also in previous investigations (cf.\ \citet{GL98,WRL13}) and is in line with experimental evidence (reviewed by \citet{EWK07}) and  theoretical predictions \citep{Gill84,Orr03}. %
Some crucial quantities related to the heuristics can be calculated explicitly in the exponential case, see Appendix~B.

\begin{figure*}[!ht]\centering
\resizebox{.63\textwidth}{!}{\input{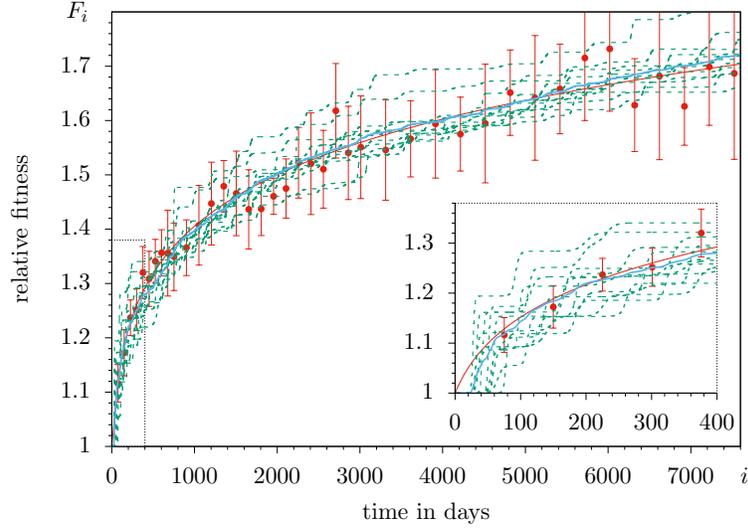}}
\caption{Simulations of the Cannings model with $X$ following $\Exp(1)$ and parameters $\widehat \varphi=0.0375$, and mutation probability $\widehat \mu=0.73$. 
}
\label{cannings_exp_init_0p15}
\end{figure*}

\begin{figure*}[!ht]\centering
\resizebox{.63\textwidth}{!}{\input{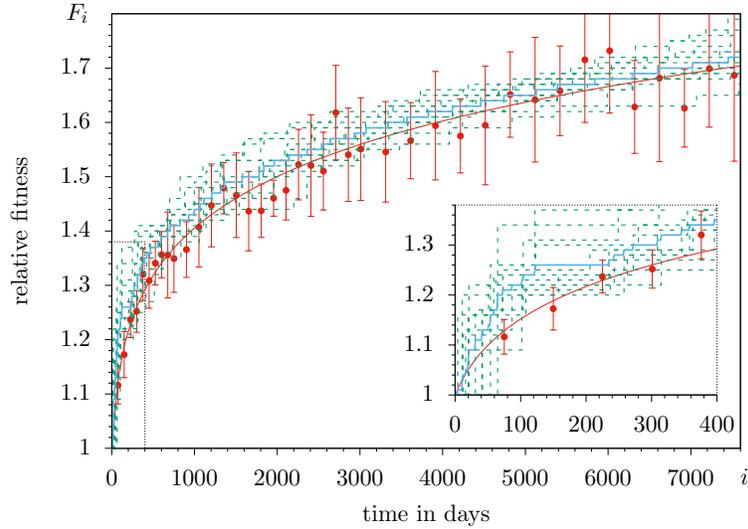}}
\caption{Simulations using the refined Gerrish-Lenski heuristics with $X$ following $\Exp(1)$ and parameters as in Fig.~\ref{cannings_exp_init_0p15}.
Mean number of clonal interference events with $x'\leqslant x$: 63; mean number of clonal interference events with $x'>x$: 29; mean number of established beneficial mutations: 19.
}
\label{approx_exp_init_0p15}
\end{figure*}

Numerical evaluation of \eqref{mu_hat} and \eqref{varphi_hat} shows
that the threshold for $\widehat{\mathfrak{d}}_1$ below which there are no solutions  $(\widehat \mu,\widehat \varphi)$  is between 0.14 and  0.15 (recall the reported value is $\widehat{\mathfrak{d}}_1=0.1$). We therefore work with $\widehat{\mathfrak{d}}_1=0.15$,
which gives $\widehat \mu=0.73$ and hence $\widehat \varphi=0.0375$  for this choice of the distribution of $X$. %
Fig.~\ref{cannings_exp_init_0p15} shows the corresponding Cannings simulations, and Fig.~\ref{approx_exp_init_0p15} displays the simulations according to the heuristics. %
The agreement of the simulation mean with the approximating power law is now nearly perfect. The fluctuations, however, are smaller in the simulations than in the experiment. As argued in Sec.~\ref{sec:GKWYLLN} in the context of the first reality check, this may be explained by the constant parameters assumed by the model, whereas parameters do vary across replicate populations in the experiment. %
\\
Let us also mention the degree of polymorphism observed in the Cannings simulations of Fig.~\ref{cannings_exp_init_0p15}. %
Counting a type as `present' if its frequency is at least 20\%, it turns out that, on average, the population is monomorphic on 79.0\% of the days; it contains two types on 19.6\% of the days, three types on 1.39\% of the days, and four or more types on 0.01\% of the days.
Thus, in the finite system, some polymorphism is present, but it is not abundant. Recall that our model does not consider neutral mutations, and thus the low level of (fitness) polymorphism observed in the simulations does not contradict the high level of genetic diversity observed in experiments \citep{Tenaillon16}.

\paragraph{Beneficial effects with a Pareto distribution.} As argued already, the exponential distribution seems to be the most realistic choice for beneficial mutation effects. %
The theory developed above, however, holds for arbitrary probability distributions on the positive half axis that have expectation 1 and a finite second moment. %
Furthermore, the analysis of the heuristics indicates that the results are, in fact, independent of the distribution, provided the compound parameter $\Gamma$ is interpreted in the appropriate way. %
It is therefore interesting to explore whether this conclusion may be verified by simulations. %
In order to push our conjecture to the limits, we have also carried out the program with  $X$ distributed according to a \emph{(shifted) Pareto distribution} (see \citet[Ch.~II.4]{Feller_71} or \citet[Ex.~2.19]{SO94}),
which again has expectation 1 but a variance that is much larger than that of Exp(1).  However, the numerical evaluations involved in the parameter estimation are substantially more difficult. We have therefore simplified the approximation for $u$ by working with $u(F)=(\log N)/s(F)$ in place of \eqref{u_N}. Our simulations (not shown here) reveal that the mean, according to the predictions in Sec.~\ref{sec:ci}, is still well described by the approximating power law, but the fluctuations are enhanced relative to the case of the exponential distribution and seem to be unrealistically large compared to the experiment. This is compatible with the statement at the beginning of this paragraph.
\section{Discussion}
\label{sec:discussion}
We have, so far, postponed a detailed comparison with the model and the results of \cite{WRL13}. We now have everything at hand to do so. 

\paragraph{Modelling aspects.}  Let us recall our starting point by summarising the key common points and differences between the WRL and the GKWY models. Namely, the common modelling assumptions are:
\begin{enumerate}
\item The population dynamics features periodic growth and dilutions, without deaths.
\item Neutral and deleterious mutations are ignored.
\item Beneficial mutations occur at constant rate, independently of the current fitness.
\item The current fitness affects the fitness increments epistatically, in a way that leads to  power laws of the same form for the mean relative fitness curve, namely \eqref{powerlaw} and \eqref{LLN}.
\end{enumerate}
The key differences are:
\begin{enumerate}
\item The GKWY model explicitly includes the runtime effect.
\item The GKWY model assumes deterministic fitness increments and ignores clonal interference, while the WRL model assumes exponentially distributed fitness increments and accounts for  clonal interference.
\item The intraday dynamics of the GKWY model is  a stochastic Yule process of growth, while deterministic synchronous divisions take place in the WRL model.
\item The fitness curve of the GKWY model results from a law of large numbers obtained via rigorous analysis, while the derivation is heuristic in the case of the WRL model. 
\end{enumerate}
In this article, we have  developed the GKWY model further  by introducing arbitrary distributions of fitness effects and taking clonal interference into account, while still obtaining a power law fitness curve. Let us now discuss the differences in detail, along with the   consequences for the interpretation of the parameters.
Here and below we use a tilde to distinguish the quantities belonging to the WRL model from our  corresponding quantities.

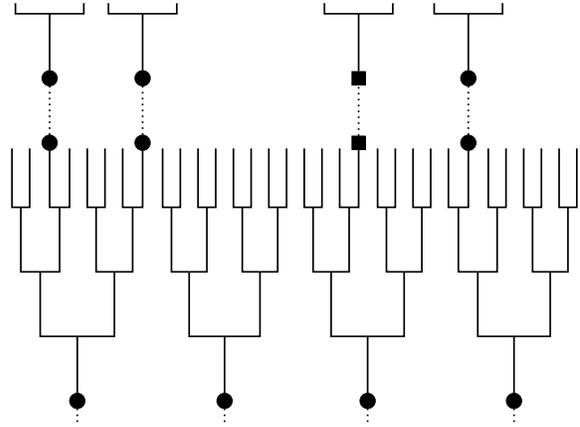
\begin{figure}
\centering
\resizebox{\columnwidth}{!}{% Need to load packages 'tikz' and 'forest' in main file!!!
%
%\usepackage{tikz}
%\usepackage{forest}
%

\forestset{
  mytree/.style={
    for tree={
      edge+={thick},
      edge path'={
        (!u.parent anchor) -| (.child anchor)
      },
      grow=north,
      scale=0.7,
      parent anchor=children,
      child anchor=parent,
      anchor=base,
      l sep=5pt,
      s sep=3pt,
      if n children=0{align=center, base=bottom}{coordinate}
    }
  }
}

\begin{forest}
mytree
% right most tree
[,phantom, edge=dotted, 
 [, edge=dotted
  [, l = 10pt, edge=dotted, draw, circle, fill=black
    [
      [
        [
          [,tier=word]
          [,tier=word]
        ]
        [
          [,tier=word]
	      [,tier=word]
	    ]
      ]
      [
        [
          [,tier=word]
          [,tier=word]
        ]
        [
          [,tier=word, draw, circle, fill=black
            [, edge=dotted, draw, circle, fill=black
              [,s sep = 27pt
               [,l=1pt]
               [,l=1pt]
              ]
            ]
          ]
          [,tier=word]
        ]
      ]
    ]
  ]
 ]
% second right most tree 
 [, edge=dotted
  [, l = 10pt, edge=dotted, draw, circle, fill=black
    [
      [
        [
          [,tier=word]
          [,tier=word]
        ]
        [
          [,tier=word]
	      [,tier=word]
	    ]
      ]
      [
        [
          [,tier=word, draw, rectangle, minimum width=9pt, minimum height=9pt, fill=black
            [, edge=dotted, draw, rectangle, minimum width=9pt, minimum height=9pt, fill=black
              [, s sep=27pt
              	[, l = 1pt]
              	[, l = 1pt]
              ]
            ]
          ]
          [,tier=word]
        ]
        [
          [,tier=word]
          [,tier=word]
        ]
      ]
    ]
  ]
 ]
% second left most tree
 [, edge=dotted
  [, l = 10pt, edge=dotted, draw, circle, fill=black
    [
      [
        [
          [,tier=word]
          [,tier=word]
        ]
        [
          [,tier=word]
	      [,tier=word]
	    ]
      ]
      [
        [
          [,tier=word]
          [,tier=word]
        ]
        [
          [,tier=word]
          [,tier=word]
        ]
      ]
    ]
  ]
 ]
%left most tree
 [, edge=dotted
  [, l = 10pt, edge=dotted, draw, circle, fill=black
    [
      [
        [
          [,tier=word, draw, circle, fill=black
            [, edge=dotted, draw, circle, fill=black
              [,s sep = 27pt
               [,l=1pt]
               [,l=1pt]
              ]
            ]
          ]
          [,tier=word]
        ]
        [
          [,tier=word]
	      [,tier=word]
	    ]
      ]
      [
        [
          [,tier=word]
          [,tier=word, draw, circle, fill=black
            [, edge=dotted, draw, circle, fill=black
              [,s sep = 27pt
               [,l=1pt]
               [,l=1pt]
              ]
            ]
          ]
        ]
        [
          [,tier=word]
          [,tier=word]
        ]
      ]
    ]
  ]
 ]
]
\end{forest}}
\caption{Synchronous growth model as used in \cite{GL98}, with equally-sized clones at the end of the day (here, $\gamma=8$); compare Fig.~\ref{fig:forest}.}
\label{fig:forest_deterministic}
\end{figure}

The main difference is that \cite{WRL13} 
%use a Cannings model to 
describe the experiment  with a discrete generation scheme given by $\log_2 \gamma \ (\approx 6.6)$ doublings during one daily growth phase, see Fig.~\ref{fig:forest_deterministic}. This neglects the variability that comes from a continuous-time intraday reproduction mechanism, and affects the WRL analogue to our formula \eqref{pi_N} for the probability to survive drift. The latter is stated in (S1) of their Supplementary Text, reads 
\begin{equation}\label{WRLs}
\widetilde \pi = \widetilde \pi(\widetilde s) = 4 \, \widetilde s,
\end{equation}
and   relies on \cite{GL98}, Appendix 1. 
In line with the generation scheme of Fig.~\ref{fig:forest_deterministic}, $\widetilde s$ is the selective advantage in each of the $\log_2 \gamma$ generations per day. At the end of the day,  the population has increased from size $N$ to size $\gamma \, N$ and consists of $N$ clones, each of (deterministic) size $\gamma$. A sampling of $N$ individuals without  replacement  thus leads to a pair coalescence probability of $(\gamma-1) / (\gamma \, N)$, and hence to an offspring variance per day of
\begin{equation}\label{WRLvariance}
\widetilde v \sim  \frac {\gamma-1}{\gamma};
\end{equation}
note the factor of 2 between $\widetilde v$ and our $v$ in \eqref{ourv}, which comes from a size-biasing effect due to the sampling from clones of random size.

Since $\widetilde s$  is related to one `doubling generation',  the selective advantage \emph{per day} is
\begin{equation}\label{sday}
\widetilde s_{\text{d}} \approx \widetilde s \, \log_2 \gamma.
\end{equation}
Now, Haldane's formula \eqref{Haldane} related to the daily rhythm gives
\[ \widetilde \pi \approx \frac{\widetilde s_{\text{d}}}{\widetilde v_{\text{d}}/2}, \]
and, with \eqref{WRLs}, this yields a  per-day offspring variance $\widetilde v_{\text{d}} \approx \log_2 \gamma$, which differs significantly  from $\widetilde v$ in~\eqref{WRLvariance} for $\gamma=100$ (to be precise, we then have $\widetilde v \approx 99/100=0.99$, whereas $\widetilde v_{\text{d}} \approx 6.6$).  Thus, we see that the ansatz of \cite{WRL13} combined with \cite{GL98} leads to an ambiguously 
defined offspring variance per day.

Moreover,  at the end of the Materials and Methods section in the Supplement, \cite{WRL13} relate the difference between the new and the old relative fitness to the (per generation) selective advantage of a mutant as follows: 
\begin{equation} \label{diffw}
w_{\text{new}} = w(1+\widetilde s)
\end{equation}
with  $\widetilde s$ from \eqref{WRLs}. Here 
\begin{equation}\label{wrecursion}
w=w_i= \frac {\log \widetilde a}{\log \widetilde b},
\end{equation}
 with the growth factors $\widetilde a = y_i(\widetilde \sigma_i)/y_i(0)$ and $\widetilde b= y_0(\widetilde \sigma_i)/y_0(0)$  as in \eqref{hatFi}.  They are not explicit about an intraday growth model, so one should think of $y_i(0), y_0(0), y_i(\widetilde \sigma_i) $ and   $y_0(\widetilde \sigma_i)$ as the numbers of individuals at the beginning and the end of the competition experiment. For a consistent definition of  the selective advantage per day, it is inevitable to use the growth factors $a_{\text{new}}$ and $a$ related to one day; then, according to \eqref{s}, one  has
\begin{equation} \label{sda}
s_{\text{d}} = \frac{a_{\text{new}}-a}a \sim \log \frac{a_{\text{new}}}a.
\end{equation}
 In principle, $a$ may (and will)  differ from the $\widetilde a$  in the definition of $w$. At least in a model with intraday exponential growth, however, the definition of $w$ in \eqref{wrecursion} becomes independent of $\widetilde \sigma_i$ because $\widetilde \sigma_i$ cancels out (see the explanation below \eqref{hatFi});  we may (and will)  therefore  use the growth factors $a=y_i(\sigma_i)/y_i(0)$ and $ b= y_0(\sigma_i)/y_0(0)$ instead of $\widetilde a$ and $\widetilde b$ in \eqref{wrecursion}. Then \eqref{wrecursion} implies
 \begin{equation} 
  \frac{w_{\text{new}}} w = \frac{1}{\log a} \Big (\log\Big (\frac {a_{\text{new}}}{a}\Big )+\log a\Big ) ,
 \end{equation}
 which by \eqref{sda} yields 
  \begin{equation} \label{wneww}
   w_{\text{new}} = w \Big (1 + \frac {s_{\text{d}}}{ \log a}\Big ),
 \end{equation}
 or equivalently, using \eqref{wrecursion} again,
\begin{equation}\label{wnewww}
 w_{\text{new}} - w =   \frac {s_{\text{d}}}{ \log b}.
 \end{equation}
Under the assumption of an intraday exponential growth we have (as long as the populations are nearly homogeneous):
\begin{equation}\label{expon}
a \approx \ee^{r\sigma},\quad b \approx \ee^\sigma, \quad w\approx r, \quad r \, \sigma \approx \log \gamma.
\end{equation}
Thus \eqref{wnewww} translates into
\begin{equation}\label{ourr}
s_{\text{d}} \approx \frac{1}{r} \,  (r_{\text{new}}-r) \log \gamma,
\end{equation}
which also results from combining \eqref{sigma} and \eqref{s_N} and equating $F$ and $r$. This shows that the runtime effect  discussed in Sec.~\ref{sec:GKWYLLN} is already implicit in the definition \eqref{wrecursion} of $w$ as the ratio of logarithms of growth factors, as soon as one uses a model with intraday exponential growth. Let
us emphasise again that this runtime effect is a consequence of the design of
Lenski's experiment; it would be absent in a variant of the experiment
in which sampling occurs at a given fixed time before the onset of the
stationary phase. Let us also note that the runtime effect appears as soon as  individuals consume the resources faster, regardless of how they developed the ability to reproduce faster. For this reason the runtime effect may play a role (and should be taken into account) in sequential dilution experiments, regardless of whether the population is monomorphic or polymorphic.
 
Furthermore, comparing \eqref{wneww} with \eqref{diffw} and using \eqref{expon} gives
\[
s_{\text{d}}  = \widetilde s \, \log a \approx \widetilde s \, \log \gamma.
\] 
Comparing with \eqref{sday}, this shows that
\[
 s_{\text{d}} = \frac{\log \gamma}{\log_2 \gamma} \, \widetilde s_{\text{d}},
\]
which points to a certain inconsistency inherent in $\widetilde s_{\text{d}}$.

Another  issue worth to compare is the interpretation of {\em diminishing returns epistasis}, and the corresponding translation between the exponent $g$ in the WRL model and the exponent $q$ in ours. Formula (S1) of \cite{WRL13}  says that the multiplicative effect on $r$ has expected size  $1/\alpha$; this corresponds to an additive effect on $r$ of expected size $
\delta:= r/\alpha$.
Thus, the ansatz \eqref{delta_N} translates into 
\begin{equation*}\frac 1\alpha = \frac \varphi {r^{q+1}}.
\end{equation*}
On the other hand, formula (S9) in \cite{WRL13} says that
\begin{equation*}\alpha = c \, \ee^{g\log r},
\end{equation*}
which implies that $g=q+1$.
%As a matter of fact, undamped multiplicative effects (which correspond to $g=0$ in \cite{WRL13}, see their formula (S6)) lead to an exponential  growth of the relative fitness (see their (S8); the fact that this growth is not super-exponential is due to the runtime effect). Note also that the power law approximation (S16) is only defined for $g>0$. 
The choice $g=1$ in the WRL model (or equivalently, $q=0$ in ours, cf.\  \eqref {powerlaw} and \eqref{LLN}) corresponds to {\em additive} increments on the Malthusian fitness that do not depend on the current value of the latter, see \eqref{delta_N}. 
%Indeed, (S9) shows that, for $g=1$, the multiplicative increment $1/\alpha$ is proportional to $1/F$ and hence corresponds to a constant {\em additive} increment. 
It is this case of constant additive increments which  may be appropriately addressed as the {\em absence of epistasis}. More precisely, in \emph{continuous time} (as considered here for the intraday dynamics), additive fitness increments correspond to independent action of mutations and hence to absence of epistasis (\citet{Fisher18}; \citet[pp.~48 and 74]{Bu00}); in \emph{discrete time}, the same would be true of multiplicative increments.  Consequently, $q= g-1$ can be seen as an exponent describing the effect of epistasis.  
With this interpretation, a (slight) concavity of the mean fitness curve is caused by the runtime effect (and hence by the design of the experiment) even in the absence of epistasis, cf.~Fig.~\ref{fig:WRL_data}.
This fact is sometimes overlooked when interpreting the mean fitness curve; see, for example, \cite{KTP09,GoDe15}. In contrast, a runtime effect closely related to ours was described by \citet{YD13} in the context of selection in fluctuating environments. Here it was observed that  selection is biased in favor of the  rare competitor, because ``more time is spent growing in environments favorable to it (because the common competitor grows more slowly, taking longer to exhaust the limiting resource)''.

A substantial part of the derivations of \cite{WRL13} deals with incorporating the Gerrish-Lenski heuristics for {\em clonal interference} into their model.
The fact that they work with
multiplicative fitness increments and various approximations complicates
the translation between the time-scaling constant in their power law
(S16) (that we subsume as $\beta$ in \eqref{powerlaw}) and our
time-scaling constant $\Gamma$ (see \eqref{LLN} and \eqref{compound_stoch}).
 We refrain from pursuing the details here; but let us  emphasise that \eqref{X} together with the calibrations discussed in Sec. \ref{sec:stoch} applies to arbitrary random (additive) fitness effects with finite second moments.

\paragraph{Analytic and simulation results.} We have presented three lines of results. First,  rigorous results for the relative mean fitness in terms of a law of large numbers in the limit $N \to \infty$ for deterministic beneficial effects in a regime of weak mutation and moderately strong selection.  Second, we have  derived transparent analytic expressions for the \emph{expected} mean fitness in a finite-$N$ system by means of  heuristics of Gerrish-Lenski type and a moment closure approximation (which is also used  by \cite{WRL13}). The beneficial effects may be either deterministic (and then require a specific thinning heuristics), or random with an arbitrary density. In the latter case we have developed a refinement of the original Gerrish-Lenski heuristics. Briefly stated, this refinement does not only consider the retainment factor \eqref{future} coming from {\em future} interfering mutations, but also the retainment factor \eqref{past} coming from {\em past} ones. This makes the heuristics consistent with its verbal description, which says that `if two contending mutations appear within the time required to become dominant in the population, then the fitter one wins.' A refinement that also includes thinning due to past competitors was suggested by \cite{Ge01};  this focuses on the \emph{time} at which a `winning' mutation appears, whereas our analysis is mainly concerned with the distribution of the \emph{effects} of these mutations. 

For reasons of calibration, we have  established an approximate analytic expression \eqref{X:delta1} for the expected scaled effect of the first beneficial mutation that goes to fixation. This introduces a \emph{size bias} into the  distribution of beneficial effects (see \eqref{scaledeffectfix}), similar to the descriptions by \citet{RVG02} and \citet{WRL13} in the case of the exponential distribution.

  As it turned out, the analytic expressions are \emph{robust}.  In particular, the estimate of $q$ is not affected by the choice of the distribution of beneficial effects, and it is also, at least approximately, independent of clonal interference, as obvious  from the independence of $F$ of the  factors $\vartheta$, $\back{\psi}$, and $\forw{\psi}$ in \eqref{constant_theta}, \eqref{psipast}, and \eqref{psifuture}. What changes is the internal structure of the compound parameter $\Gamma$, but for any given estimate $\widehat \Gamma$, the mutation probability and  scaling of beneficial effects may be arranged appropriately (provided $X$ has second moments). The deviations from $q=0$ are a signal of diminishing returns epistasis; at this point, let us  emphasise again that the approximating curve of the mean relative fitness is (slightly) concave even for $q=0$ (due to the runtime effect). By any means, the pronounced concavity in the  curve approximating the LTEE data  (with its estimated $\widehat q = 4.2$) gives strong evidence for diminishing returns epistasis, in line with the conclusions of previous investigations \citep{WRL13,GoDe15,Wuetal17}. We would like to emphasise, however, that our goal here was not to find the `best' (or even the `true') increment function; rather, the choice \eqref{delta_N} was made for the sake of comparison with \cite{WRL13}, while we have seen that the GKWY model in fact allows for arbitrary increment functions \eqref{X:delta_general}.

Our third line of investigations is a simulation study both of the Cannings model and the approximating heuristics (described in Section \ref{sec:det} for deterministic effects and in Section \ref{sec:stoch} for stochastic effects). It turned out that the heuristics (which might be improved even further by taking into account the refined heuristics of \citet{Ge01} and \citet{RVG02}) approximates the Cannings model quite well.  The simulations show that the deviation of the mean fitness of the latter from the power law fitted to empirical data is moderate for deterministic increments and minute for exponential increments.

\paragraph{Validity and limits of the  sequential fixation model.} Following \cite{WRL13}, we have worked with a sequential fixation model, which led to the dynamical law of large numbers of \cite{GKWY16} in the limit $N \to \infty$ under scaling assumptions on $\mu$ and $\varphi$ that, in particular, require $\mu \ll \varphi$ and $\mu, \varphi \to 0$ as $N \to \infty$. It then turned out that, provided appropriate corrections for clonal interference are made, the power law  still describes the mean of the simulations very well for finite $N$ and  moderately-large $\mu$ and $\varphi$, even when $\mu$ is substantially larger than $\varphi$. In this sense, the result once more appears to be quite robust.

A question that still remains concerns the `true' beneficial mutation probability and the `true' distribution of the beneficial effects. In particular, it is not finally decided (neither by experiment nor by theory) whether the fitness trajectory  increasing from 1 to $\approx 1.7$ is dominated by a small number of mutations of large effects or a larger number of smaller effects. On the one hand,  the reported mean fitness increment of the first fixed beneficial mutation, $\widehat{\mathfrak{d}}_1=0.1$, is quite large, and if this is taken as typical, it is hard to reconcile with a plethora of small effects. On the other hand, \cite{Tenaillon16} infer that most of their 60--110 fixed mutations are beneficial in those populations that keep the original low mutation rate; whereas the adaptive proportion is harder to quantify in those strains that evolved into hypermutators. Our 27 or 19 fixed beneficial mutations (for deterministic and exponential effects, respectively), as estimated from the  trajectory simulated according to the heuristics and averaged over \emph{all} populations, may seem somewhat on the low side, but this may also reflect the fact that the parameters are close to the limit of validity of a sequential fixation model with clonal interference heuristics, as discussed in Secs.~\ref{sec:ci} and \ref{sec:stoch}.

\section*{Appendix~A: Derivation of effect of first fixed mutation}

We derive, in our Poisson model, a system of equations that relates $\mu$ and $\varphi$ to the observable quantities $\Gamma$ and $\mathfrak{d}_1$, where $\mathfrak{d}_1$ is the \emph{expected scaled effect of the first among the contending mutations} (in fitness background $F=1$) \emph{that is not killed}. 
To this end, we consider a sequence of points $(\mathcal T_j, X_j)_{j\geqslant1}$ in $\Pi$ (the Poisson process of contending mutations) that is strictly monotonic increasing in both coordinates and inductively defined as follows. %
\\
$(\mathcal T_1, X_1)$ is the point in $\Pi$ with the smallest $\tau$-coordinate, and given $(\mathcal T_{j}, X_{j})$, 
$(\mathcal T_{j+1}, X_{j+1})$ is the point in $\Pi$ %which, among all points with  $x$-coordinate larger than $X_j$ and $\tau$-coordinate larger than $\mathcal T_i$, is the one with the smallest \mbox{$\tau$-coordinate}.
with
\[
\mathcal T_{j + 1} = \min \{ \tau: (\tau, x) \in \Pi, \, \tau > \mathcal T_{j},\,  x > X_j \}.
\]
Again we say that $(\mathcal T_{j+1}, X_{j+1})$ kills $(\mathcal T_{j}, X_{j})$, if $\mathcal T_{j+1} < \mathcal T_j + u(X_j)$.%
\footnote{As long as we assume $F = 1$, we suppress the first argument in the functions defined in \eqref{X} for notational convenience.} %
Let 
\[
Z:= \min\{j \geqslant 1: \mathcal T_{j+1} > \mathcal T_j + u(X_j)\},
\]
i.e. $(\mathcal T_Z, X_Z)$ is the earliest among the points $(\mathcal T_j,X_j)$, $j=1,2,\ldots$, that is \emph{not killed}. %
The point $(\mathcal T_Z, X_Z)$ is thus called the (first) \emph{winner}; at time $\mathcal T_Z$, the relative fitness of the population jumps from 1 to $1+\varphi \,  X_Z$. %
\\
Our aim is to find  the density of the \mbox{$x$-coordinate} of the winner,
\begin{linenomath}
\begin{equation}\label{XZ}
%w(x)\, dx := 
w(x) \D{x} = \mathbb P(X_Z \in \D{x}), \quad x \geqslant 0.
\end{equation}
\end{linenomath}
%\todo[inline]{explain; Ref.1 and 4}
%
Note that \eqref{XZ} (as well as other formulae below) is a convenient differential notation which gets its (unambiguous) meaning under the integral.
From elementary properties of Poisson processes we infer that, given %$(\mathcal T_1,X_1), \ldots, 
$(\mathcal T_{j},X_{j})$, the waiting time $\mathcal T_{j+1}-\mathcal T_j$ is exponentially distributed with parameter 
\begin{linenomath}
\begin{equation*}
\mu \int_{X_j}^\infty \pi(y) \, h(y) \D{y}.
\end{equation*}
\end{linenomath}
Hence, with $\chi(x):= \forw{\chi}(1, x)$ from \eqref{future}, 
\begin{linenomath}
\begin{align*}
\chi(y) 
&= \mathbb P(\mathcal T_{j+1}-\mathcal T_j > u(y) \mid X_j=y), 
\intertext{which, in turn,}
&= \mathbb P(Z=j \mid Z\geqslant j, \, X_j=y),
\end{align*}
\end{linenomath}
the conditional probability that the $j$-th of the increasing points is the winner, given that all the previous ones have been killed. %
Moreover, given %$(\mathcal T_1,X_1), \ldots, 
$(\mathcal T_{j},X_{j})$, the random variables $\mathcal T_{j+1}-\mathcal T_j$ and $X_{j+1}$ are independent, and $X_{j+1}$ has the conditional density 
\begin{linenomath}
\begin{equation}
\label{Xnext}
\begin{split}
 \mathbb P(& X_{j+1} \in \D{x} \mid X_j=y) \\
 &= \frac{\pi(x) \, h(x) \D{x}} { \int_y^\infty \pi(y') \, h(y') \D{y'}} =:\rho(x \mid y) \D{x}
\end{split}
\end{equation}
\end{linenomath}
for $x \geqslant y \geqslant 0$. % 
Consequently, the conditional density to be affected by the next killer in $x$ is
\begin{equation*}
\begin{split}
 \mathbb P(Z \geqslant j+1, X_{j+1} \in \D{x} \mid Z\geqslant j,\,  X_j=y) \\ 
 = (1-\chi(y)) \, \rho(x \mid y) \D{x}.
\end{split}
\end{equation*}
With $x_0:=0$, this gives the following formula for the joint distibution of $X_1,\ldots, X_Z$ and $Z$: %
\begin{linenomath}\postdisplaypenalty=0
\begin{align} 
\mathbb P( X_1 & \in  \D{x_1}, \ldots, X_{k-1}\in \D{x_{k-1}}, X_k\in \D{x}, Z=k)\notag\\
 & = \prod_{j=1}^{k-1} \big[\rho(x_{j} \mid x_{j-1}) \, (1- \chi(x_j)) \big]  \label{well}
 \\  &  \quad \cdot \rho(x \mid x_{k-1}) \, \chi(x) \D{x_{1}}\cdots \D{x_{k-1}}\D{x}. \notag
 \end{align}
%where $ B_{\ell-1}&(x_1, \ldots, x_{\ell-1})$
%\begin{align*}
%  \phantom{AAAAA} = \prod_{j=1}^{\ell-1} \big[\rho(x_{j} \mid x_{j-1}) \, (1- \chi(x_j)) \big].
%\label{product}
%\end{align}
\end{linenomath}
The density \mbox{$w_k(x) \D{x} = \mathbb P(X_Z \in  \D{x}, Z  =   k)$}
%arises by integrating \eqref{well} over $x_1, \ldots, x_{k-1}$, under the constraints $0\leqslant x_1\leqslant \cdots \leqslant x_{k-1}\leqslant x$. %
arises by integrating \eqref{well} over $\mathcal{F}_k(x):= \{x_1, \ldots, x_{k-1} \mid 0\leqslant x_1\leqslant \cdots \leqslant x_{k-1}\leqslant x\}$, where $\mathcal{F}_1 := \varnothing$. %
We then sum over all $k \leqslant \ell$ and, with the help of
\eqref{Xnext}, obtain 
\begin{linenomath}\postdisplaypenalty=0
\begin{align}
\mathbb P(X_Z \in \D{x}, Z & \leqslant \ell) = \sum_{k = 1}^\ell \mathbb P(X_Z \in \D{x}, Z = k)  \notag\\
& = \,  \pi(x)  \, h(x)\, \chi_\ell^{}(x) \D{x}, \label{glell}
\end{align}
\end{linenomath}
where  
%\[ 
%\begin{split}
%& \chi_\ell^{}(x) = \chi(x)+ \sum_{k = 2}^\ell \frac{1}{\int_{x_{k-1}}^\infty \pi(y)\,  h(y) \D{y}} \\
% & \times \int_0^x \int_{x^{}_1}^x\dots \int_{x^{}_{k-2}}^x 
%  \prod_{j=1}^{k-1} \big[\rho(x_{j} \mid x_{j-1}) \, (1- \chi(x_j)) \big]  \\
%   & \textcolor{white}{blah}\times \D{x_{k-1}} \cdots \D{x_2} \D{x_1} \, \chi(x)
%\end{split}
%\]
\[ 
\begin{split}
& \chi_\ell^{}(x) = \chi(x) \sum_{k = 1}^\ell \frac{1}{\int_{x_{k-1}}^\infty \pi(y)\,  h(y) \D{y}}  \\
 & \times \! \! \! \int\limits_{\mathcal{F}_k(x)}
  \prod_{j=1}^{k-1} \rho(x_{j} \mid x_{j-1}) \, (1- \chi(x_j)) \D{x_{k-1}} \cdots \D{x_1}
\end{split}
\]
(with the multiple integral  $:=1$ for $k = 1$)
plays the role of an additional reweighting factor and coincides with \eqref{future} in the case $\ell =1$, normalised by $\mathbb{E}(\pi(X))$. %

Then, for the density of $X_Z$ \emph{conditional on} $Z\leqslant \ell$, we obtain %
\begin{equation}
%w_\ell(x) := 
%\mathbb P(X_Z \in \D{x} \mid \,  Z \leqslant \ell) \approx \frac{ \density(x) \, x \,\psi_\ell(\mu,x) \D{x}}{\EE\big (\psi_\ell(\mu,X) X \big )} 
\begin{split}
\mathbb P(X_Z \in \D{x} \mid & Z \leqslant \ell) = \\
& \frac{  \pi(x) \,\chi_\ell(x) \, h(x)  \D{x}}{\int_0^\infty \pi(x)\, \chi_\ell(x) \, h(x)  \D{x}}, 
\end{split}
\label{scaledeffectfix}
\end{equation}
%
%as an approximation 
which should be very close to \eqref{XZ} provided the number of contending mutations present simultaneously is rarely larger than $\ell$. %
\\
Consequently, with the approximations \eqref{X} as well as \eqref{psifuture}, and $\psi_\ell$ taking the place of $\chi_\ell^{}$ in the approximate analogue of \eqref{glell},
\begin{equation}
\begin{split}
 	\EE\big ( \delta(X_Z) \mid Z\leqslant \ell \big ) = 	\varphi \, \EE\big ( & X_Z \mid Z \leqslant \ell \big ) \\
  	\approx\varphi\, \frac{\EE\big (\psi_\ell (X) \, X^2 \big )}{\EE\big (\psi_\ell(X) \, X\big )} 
  	& =: \varphi \, \zeta_\ell^{}(\mu, \varphi).
 \end{split}
 \label{X:delta1}
\end{equation}
Note that, under the assumptions on $X$, %the right-hand side of \eqref{X:delta1} 
$\zeta_\ell(\mu, \varphi)$ (as an approximation of the expectation of the scaled  effect of the first beneficial mutation that goes to fixation) is parametrised by $\mu$ and $\varphi$ (via $\psi_\ell$) and well defined for any $\mu$ and $\varphi$, since $0<\psi_\ell \leqslant 1$. %
\\
Then, again with the above constraint on $\ell$, the left-hand side of \eqref{X:delta1} should be a good approximation for the expected value of the mean fitness increment due to the first fixed beneficial mutation, which we denote by $\mathfrak{d}_1$. %
Then, \eqref{compound_stoch} together with \eqref{X:delta1} renders 
%two equations for $\mu$ and $\varphi$, which can be used numerically to obtain estimates $\widehat \mu$ and $\widehat \varphi$. 
the system of equations \eqref{mu_hat}.
%%
%\begin{subequations}
%\label{estimateparameters}
%\begin{linenomath}\postdisplaypenalty=0
%\begin{align}
% \frac{\mu \, I(\mu, \varphi)}{(\zeta_\ell^{}(\mu, \varphi))^2} &= \frac{\Gamma}{C \, \mathfrak{d}_1^2 \log \gamma} \label{mu_hat_app}, \\
% %\EE\big (\psi_\ell(\mu,X)X\big )^2}{\EE\big (\psi_\ell(\mu,X) X^2 \big )^2} \label{mu_hat}, \\
%\varphi &= \frac{\mathfrak{d}_1}{\zeta_\ell^{}(\mu, \varphi)}. \label{varphi_hat_app}
%\end{align}
%\end{linenomath}
%\end{subequations}
%
%where $\mu$, as the solution of \eqref{mu_hat}, determines $\varphi$ via \eqref{varphi_hat}. %

\section*{Appendix~B: Application to Exp(1)-distributed $X$}

The choice of an $\Exp(1)$-distributed $X$ is not only canonical and in line with experience, but also provides convenient explicit expressions for some crucial quantities appearing in our framework. Note first that, independently of the choice of $h$ and just by using \eqref{Xnext} together with \eqref{X:pi}, we  obtain
\[
\rho(x \mid y) \approx \frac{x\, h(x)}{\int_y^\infty y' \, h(y') \D{y'}}
\]
and, in particular,  $\rho(x \mid 0) \approx x\, h(x)$. %
Now, for  $h(x) = \ee^{-x}$, after applying \eqref{X:pi} and \eqref{X:u} to \eqref{psipast} and \eqref{psifuture}, we obtain
\begin{linenomath}\postdisplaypenalty=0
\begin{align*}
&\back{\psi} (x) = \exp\Big(\! -\mu \, C \, \big(\Gamma(0,x) + \ee^{-x}\log(N \, \varphi \, x) \big) \Big), \\
&\forw{\psi} (x) = \exp\Big(\! -\mu \, C \log(N \, \varphi \, x) \, \ee^{-x} \, \frac{1 + x}{x} \Big),  
\intertext{and}
&\bafo{\psi} (x) = \\
& \ \exp \Big(\! - \mu \, C \big(\ee^{-x} \, \log (N \, \varphi \, x) \frac{1+2x}{x} + \Gamma(0,x) \big) \Big),
\end{align*}
\end{linenomath}
where $\Gamma(0,x) = \int_x^\infty t^{-1} \ee^{-t} \D{t}$ means the incomplete gamma function. %
Then, Eq. \eqref{well} yields 
\begin{linenomath}\postdisplaypenalty=0
\begin{align*}
w_1(x) \D{x} & \approx x\, \ee^{-x} \, \forw{\psi} (x), \\
w_2(x) \D{x} & \approx w_1(x)\,\int_0^x \frac{y}{y + 1} \, \big(1 - \forw{\psi}(y) \big) \D{y}, 
\intertext{and}
w_3(x) \D{x} & \approx w_1(x)\,\int_0^x \frac{y}{y + 1} \, \big(1 - \forw{\psi}(y) \\ 
			 & \quad\quad\quad \times \int_y^x \frac{z}{z + 1} \, \big(1 - \forw{\psi}(z) \big) \D{z} \D{y}.
\end{align*}
Let us mention that similar expressions may  also be recovered for  $X$ following a  Pareto distribution (at least for all quantities that only depend on $\forw{\psi}$). %
Unfortunately, the dependence of the logarithmic term in \eqref{psipast} on the integration variable makes it much more difficult to obtain explicit expressions for $\back{\psi}$, $\bafo{\psi}$ and, finally, $I(\mu, \varphi)$ of \eqref{compound_stoch}. The latter, in turn, is required to find a solution of \eqref{system_est} (recall our approximation at the end of Sec.~\ref{sec:stoch}). 
\end{linenomath}

\section*{Appendix~C: Simulation algorithms}

%defs for algorithms
\renewcommand{\algorithmcfname}{Algorithm}
% Settings for algorithm2e
\DontPrintSemicolon % no ';' at the end of lines
\LinesNumbered      % line numbering
\SetAlgoLined       % horizontal lines
\SetAlgoVlined      % vertical lines
%

%###########
% MAKROS
%###########
% Common
\newcommand{\iter}[2]{{#1}^{(#2)}}
\newcommand{\relF}{\boldsymbol{F}}
\newcommand{\mutProb}{\widehat{\mu}}
\newcommand{\benEffect}{\widehat{\varphi}}
\newcommand{\episEffect}{\widehat{q}}
\newcommand{\param}[1]{\textcolor{red}{#1}}
\newcommand{\chose}[1]{\textcolor{blue}{#1}}

% Cannings specific
\newcommand{\F}{\mathcal{R}}
\newcommand{\N}{\mathcal{N}}
\newcommand{\NB}[2]{\text{NB}{(#1, #2)}}
\newcommand{\Poi}[1]{\text{Poi}(#1)}
\newcommand{\types}{\text{typ}}
\newcommand{\mutants}{\text{mut}}
\newcommand{\desc}{\text{des}}

% Approx specific
\newcommand{\tjumps}{\boldsymbol{\iota}}
\newcommand{\fitincr}{\delta}
\newcommand{\fitincrold}{\delta^{\,\uparrow}}
\newcommand{\tdur}{\iota}
\newcommand{\BerD}[1]{\text{Ber}(#1)}
%###########

Let us briefly describe the two algorithms we have used to simulate our model. Before we come to the details, let us say a few words about  notation and strategy. We will throughout use the framework  \eqref{X},  which reduces to  \eqref{delta_N}--\eqref{s_N}, \eqref{pi_N}, and \eqref{u_N} in the case of deterministic beneficial effects, where $X \equiv 1$, that is, the distribution of $X$ is a point measure on 1. (But note that, in Sec.~\ref{sec:stoch}, we assume  that $X$ has a density; this implies that any two realisations of $X$ are different with probability 1, so that there is a clear `winner' in the Gerrish-Lenski heuristics. The analysis of Sec.~\ref{sec:stoch} therefore does not carry over to the deterministic case.)
\emph{Curly} symbols indicate \emph{sets} of values, whereas \emph{bold} symbols indicate \emph{lists} and $\iter{\bullet}{k}$ their $k$-th element. 

\paragraph{Algorithm~\ref{alg:cannings}} performs an individual-based simulation  of the Cannings model with selection, as formulated in Section~\ref{sec:GKWYLLN}. Its iterations are based on real-world days $i$. The algorithm keeps track of the  sizes $\N_j$ of the classes (or subpopulations) of individuals that have  reproduction rate $\F_j$, $j \geqslant 1$. As long as $n_\types$, the number of different  reproduction rates in the population, equals 1, the population is homogeneous, so that  the intraday growth and subsequent sampling do not change the current state. If $n_\types > 1$,  we use the fact that the clone size at time $\sigma$ in a Yule process with branching rate $\F_j$ started by a single individual is 1 plus a random variable that follows Geo$(\ee^{-\F_j \sigma})$, the geometric distribution\footnote{We take Geo($p$) as the distribution of the numbers of failures before the first success in a coin tossing with success probability $p$.} with parameter $\ee^{-\F_j \sigma}$ (cf.\ \citet[Ch.~XVII.3]{Feller_68} or \citet[Ch.~1.3.3]{Dur08}). The size of the corresponding subpopulation at time $\sigma$ is then $\N_j$ plus the sum of $\N_j$ independent copies of the geometric random variable.  This sum follows $\NB{\N_j}{\ee^{-\F_j \sigma}}$, the negative binomial distribution with parameters $\N_j$ and $\ee^{-\F_j \sigma}$, cf.\ \citet[Ch.~VI.8]{Feller_68} or \citet[pp.~168/169]{SO94}.  The only point where each individual must be treated separately is the sampling step, where $N=5 \cdot 10^6$ new founder individuals are drawn without replacement from the $\approx 5 \cdot 10^8$ descendants. After the sampling, the number of mutation events is drawn from $\Poi{\mutProb}$, the Poisson distribution with parameter $\mutProb$ (line~\ref{alg:cannings:nmut}). The affected individuals are then chosen \emph{uniformly without replacement} from among the $N$ new founders. 

\paragraph{Algorithm~\ref{alg:approximation}} unifies the two versions of the thinning heuristics of Sec.~\ref{sec:ci}. %
We now only keep track of  mutations that effectively lead to an increase of the relative fitness, and thus have a homogeneous population in every iteration $k$. %
The number $k$ counts the fixation events and the vector $\tjumps$ ($\relF$) holds the times (relative fitness values) at which they occur. % 
More precisely, we retain the fittest contender represented by the triple $(\back{\iota}, \back{u} , \back{\delta})$ of its time of appearance together with its inherent dead time and fitness increment according to \eqref{X}. %
The fixation of the fittest contender is queried until the total waiting time $\cont{\tau}$ of appearing mutations (with increment $\cont{\delta}$) exceeds the dead time $\back{u}$. %
New mutations appear after waiting times $\Delta$ following Geo$(\mutProb)$. %
For every such mutation, it is immediately decided whether or not it survives drift by drawing a Bernoulli random variable with success probability $\pi$ according to \eqref{X:pi} (line~\ref{alg:approximation:bernoulli}).  %
If the mutation survives and carries a larger fitness increment than the fittest contender, it \emph{replaces} the fittest contender and $\cont{\tau}$ is reset (line~\ref{alg:approximation:ci}). %
If the total waiting time exceeds $\back{u}$, then the fittest (\emph{and} not yet fixed) contender goes to fixation (thus changing the population's relative fitness and its length of the growth period $\sigma$); and the triple representing the fittest contender is then reset (line \ref{alg:approximation:fixation}). %
For the choice $X \equiv 1$, this means that the first out of two competing mutations always wins; the case of \texttt{Fitter contender appeared during $\back{u}$ \& replaces $(\back{\iota}, \back{u}, \back{\delta})$} can \emph{only} occur if no fittest contender is already queried after the last fixation event.

For the sake of completeness, the parameter combinations for the simulations in this paper are summarised in Tab.~\ref{tab:summary:parameter:values}.

\begin{table*}[h]
  \centering
  \begin{tabular}{cccccccc}

    \toprule
 		Law of $X$      &  $\iota_{\max}$       & $\episEffect$  & $\big(\widehat{\mathfrak{d}}_1\big)$                &	$\mutProb$ & $\benEffect$ & Algo.~\ref{alg:cannings} & Algo.~\ref{alg:approximation}
 		\\
 		\midrule
 		$\equiv 1$    &  $7600$          & $4.20$       &  $(0.110)$        &	$0.0570$ & $0.110$ & Fig.~\ref{cannings_det_init_0p11_uncorrected} & 
 		\\
 		$\equiv 1$    &  $7600$           & $4.20 $  &  $(0.110)$              &	$0.242$  & $0.110$ & Fig.~\ref{cannings_det_init_0p11} & Fig.~\ref{approx_det_init_0p11}
 		\\
	 	Exp($1$)       &    $7600 $          & $4.20$    &  $(0.150)$                  &	$0.730$  & $0.0375$ & Fig.~\ref{cannings_exp_init_0p15} & Fig.~\ref{approx_exp_init_0p15}
	 	\\
%	 	shifted Pareto, &   \multirow{2}{*}{$7600$}  & \multirow{2}{*}{$4.2$}  & \multirow{2}{*}{$(0.3)$} &	\multirow{2}{*}{$0.14$} & \multirow{2}{*}{$0.03$} & \multirow{2}{*}{Fig.~\ref{cannings_par_alp_2p5_init_0p3}} & \multirow{2}{*}{Fig.~\ref{approx_par_alp_2p5_init_0p3}}
%	 	\\
%	 	$a = 2.5$, cf. \eqref{pareto}  &   &  &  & & 
%	 	\\
    \bottomrule
  \end{tabular}
  
  \caption[Summary of  parameter values for simulations]{Summary of parameter values for simulations. The population size and the dilution factor have been fixed as $N = 5 \cdot 10^6$ and $\gamma = 100$ throughout.}
  \label{tab:summary:parameter:values}
\end{table*}

\IncMargin{1em}
\begin{algorithm*}%[h]
  \caption[Simulating Lenski's experiment (Cannings model)]{Simulating Lenski's experiment (Cannings model)}
  \label{alg:cannings}

  % Input
  \SetKwInOut{Input}{Input}
  \Indm
  \Input{User chosen density law of $X$ and parameters $\iota_{\max}^{}$, $\episEffect$, $\mutProb$, $\benEffect$.}

  \Indp
	\textbf{Initialise} $k = 0$; $\sigma = 1$; $n_\types = 1$, $n_\mutants = 0$; $\F = \{1\}$, $\mathcal{\N} = \{N\}$.\;
	\While {$k < \iota_{\max}$}
	{
		\tcp{Length of intraday growth time}
		Solve \eqref{def_sigma}, i.e. $\sum\limits_{j = 1}^{n_\types} \N_j e^{\F_j \sigma} = \gamma N$, to obtain $\sigma$.\;
		Set $\iter{\relF}{k}$ according to \eqref{rel_fitness}.\;
		\If{$n_\types > 1$}
		{
			\tcp{Intraday population growth}
			$n_\desc \gets 0.$\;
			\For {$j = 1, \ldots, n_\types$}
			{
				Draw $D \sim \NB{\N_j}{e^{- \F_j \sigma}}$ and set $n_\desc \gets n_\desc + \N_j + D$.\;
				\label{alg:cannings:D} 
			}
			\tcp{Interday sampling}
			Draw sample $\{j_1, \ldots, j_N \}$ without replacement from $\{1, \ldots, n_\desc\}$ and set $\N = \{\N_1, \ldots, \N_{n_\types} \}$ accordingly.\;
			\For{$j = 1, \ldots, n_\types$}
			{
				\If{$\N_{j} = 0$}
				{
					Remove type $j$ and set $n_\types \gets n_\types - 1$.\;
				}
			}
		}
		\tcp{Mutation}
		Draw $n_\mutants \sim \Poi{\mutProb}$ and set $n_\types \gets n_\types + n_\mutants$.\;
		  \label{alg:cannings:nmut}
		\If{$n_\mutants > 0$}
		{
			Draw sample $\{i_1, \ldots, i_{n_\mutants}\}$ without replacement from $\{1, \ldots, N\}$.\;
			\For{$j = 1, \ldots, n_\mutants$}
			{
				$\N_{i_j} \gets \N_{i_j} - 1$ and $\N \gets \N \cup \{1\}$.\;
				Draw $X$ and set $\F \gets \F \cup \{ \F_{i_j} + \delta(\F_{i_j}, X) \}$ according to \eqref{X:delta}.\;
				\If{$\N_{i_j} = 0$}
				{
					Remove type $i_j$ and set $n_\types \gets n_\types - 1$.\;
				}
			}
		}
		$k \gets k + 1.$\;
	}
  \Return{$\relF$.}\;

\end{algorithm*}

\begin{algorithm*}[!ht]
\SetKwRepeat{Do}{do}{while}
\small
% Title
\caption[Simulating Lenski's experiment (thinning heuristics)]{Simulating Lenski's experiment (thinning heuristics)}
\label{alg:approximation}
% Input
\SetKwInOut{Input}{Input}
\Indm
\Input{User chosen law of $X$ and parameters $\iota_{\max}^{}$, $\episEffect$, $\mutProb$, $\benEffect$.}
\Indp
\textbf{Initialise} $k = 0$; $\iter{\tjumps}{k} = 0, \iter{\relF}{k} = 1$; $\back{\iota} = \back{u} = \back{\delta} = 0$; $\cont{\tau} = \cont{\delta} = 0$; $\sigma = \sigma(\iter{F}{k})$.\;
\While {not terminated, i.e. $\back{\iota}  \leq \iota_{\max} \wedge k \leq k_{\max}$}
{
\Do {$S$ following $\BerD{ C \, \cont{\delta} \, \sigma}$ is unsuccessful according to \eqref{X:pi}	\label{alg:approximation:bernoulli}}
{
Draw $\Delta$ following Geo($\mutProb$) and set $\cont{\tau} \gets \cont{\tau} + \Delta$.\;
\If{Fittest contender $(\back{\iota}, \back{u}, \back{\delta})$ is not fixed yet}
{
\If{$\cont{\tau} > \back{u}$}
{
\tcp{Fixate fittest contender $(\back{\iota}, \back{u}, \back{\delta})$.}
\label{alg:approximation:fixation}
$(\iter{\tjumps}{k + 1}, \iter{\relF}{k + 1}) \gets (\back{\iota}, \iter{\relF}{k} + \back{\delta} )$\;
$k \gets k + 1$\;
$(\back{\iota}, \back{u}, \back{\delta}) \gets (\iter{\tjumps}{k},0,0)$\;
$\sigma \gets \sigma(\iter{\relF}{k})$\;
}
}
Draw $X$ and set $\cont{\delta} \gets \benEffect \, X \, (\iter{\relF}{k})^{ - \episEffect}$ according to \eqref{X:delta}.\;
}
\If{$\cont{\delta} > \back{\delta}$}
{
\tcp{Fitter contender appeared during $\back{u}$ \& replaces $(\back{\iota}, \back{u}, \back{\delta})$}
\label{alg:approximation:ci}
$(\back{\iota}, \back{u}, \back{\delta}) \gets (\back{\iota} + \cont{\tau}, \log(C \, \cont{\delta} \, (\iter{\relF}{k})^{\episEffect}) / (\cont{\delta} \, \sigma), \cont{\delta})$ according to \eqref{X:u}. \;
$\cont{\tau} \gets 0$.
}
}
\Return{$(\tjumps, \relF)$.}\;
\end{algorithm*}
\clearpage

\subsection*{Acknowledgements}
It is our pleasure to thank Phil Gerrish for valuable hints and comments on the manuscript. The paper also profited from discussions with Jason Schweinsberg about the WRL model and its analysis; he further  pointed us to a strategy to  relax the lower bound on the order of the selection strength in \citet{GKWY16}, see the discussion in the paragraph {\em Scaling regime and law of large numbers} in Sec. 2.
Furthermore, we thank Richard Lenski for stimulating discussions and Nick Barton for sharing with us his thoughts about~\eqref{u_N}.  Last not least, we are indebted to four anonymous referees, who provided numerous valuable hints to improve the manuscript. This project received financial support from Deutsche Forschungsgemeinschaft (German Research Foundation, DFG) via Priority Programme SPP 1590 \emph{Probabilistic Structures in Evolution}, grants no. BA 2469/5-2 and WA 967/4-2.

\bibliographystyle{plainnat}

\end{document}